\documentclass[aps, prd, preprint, groupedaddress,showpacs,18pt]{revtex4}
\usepackage{hyperref}
\usepackage{mathrsfs}
\usepackage{bbm}
\usepackage{amsfonts}
\usepackage{amsmath}
\usepackage{graphicx}
\usepackage{subfig}
\usepackage{feynmp}

\newcommand{\void}[1]{}

\allowdisplaybreaks{}

\def\bea{\begin{eqnarray}}
\def\eea{\end{eqnarray}}
\def\beq{\begin{equation}}
\def\eeq{\end{equation}}
\def\to{\rightarrow}
\newcommand{\beas}{\begin{eqnarray*}}
\newcommand{\eeas}{\end{eqnarray*}}
\newcommand{\ba}{\begin{array}}
\newcommand{\ea}{\end{array}}

\def\be{\begin{equation}}
\def\ee{\end{equation}}
\def\bea{\begin{eqnarray}}
\def\eea{\end{eqnarray}}
\def\lmd{\lambda}

\newcommand{\td}[1]{\tilde{#1}}

\newcommand{\bary}{\begin{array}}
\newcommand{\eary}{\end{array}}

\newcommand{\ke}{k_{\mbox{\tiny off}}}
\newcommand{\Le}{L_{\mbox{\tiny off}}}
\newcommand{\Ve}{V_{\mbox{\tiny off}}}
\def\nb{\nonumber}

\begin{document}

   \title{Ward Identity Implies Recursion Relation at Tree and Loop Level }

\author{Yun Zhang\footnote{zyzhangyun2003@gmail.com} and Gang Chen\footnote{corresponding author: gang.chern@gmail.com}}
{\affiliation{Department of Physics, Nanjing University\\
22 Hankou Road, Nanjing 210093, China
}

\hspace{1cm}
\begin{abstract}
In this article, we use Ward identity to calculate tree and one loop level off shell amplitudes in pure Yang-Mills theory with a pair of external lines complexified. We explicitly prove Ward identity at tree and one loop level using Feynman rules, and then give recursion relations for the off shell amplitudes. We find that the cancellation details in the proof of Ward identity simplifies our derivation of the recursion relations. Then we calculate three and four point one loop off shell amplitudes as examples of our method.
\end{abstract}


\date{\today}
\maketitle

%about two off shell tree level amplitudes

\section{Introduction}

%\textcolor{red}{add some analysis of the off shell amplitudes in case of collinear limit etc..?}%should we mention the importance or usage of off-shell amplitudes

At tree level, the amplitudes of pure Yang-Mills fields can be written as rational functions of external momenta and polarization vectors in spinor form \cite{Parke:1986gb,Xu:1986xb,Berends:1987me,Kosower,Dixon1,Witten1}. Such rational functions can be analyzed in detail in algebra system. According to this, BCFW recursion relation was proposed and developed in \cite{Britto:2004nj,Britto:2004nc,Britto:2004ap}, and then proved in \cite{Britto:2005fq} using the pole structure of the tree level on shell amplitudes. This has been an exiting progress on the amplitudes in pure Yang-Mills theory. For theory with massive fields \cite{Badger1,Ozeren,Schwinn,Chen1,Chen2}, the amplitudes are also rational functions of external momenta and polarization vectors in spinor form.

%For theory with massive fields, the amplitudes are also rational functions of external momenta and polarization vectors in spinor form, and there are progresses in this direction \cite{Badger1,Ozeren,Schwinn,Chen1,Chen2}.   

At loop level, although the whole amplitudes are no longer rational functions in general, they can be decomposed into some basic scalar integrals with coefficients being rational functions of external spinors \cite{BernD1,BernD2}. The coefficient structures are studied in depth in \cite{Dixon4,Bern,Bern1}. On the other hand, the integrands of the amplitudes are rational functions of the external spinors and integral momenta. For the N=4 planar super Yang-Mills theory, \cite{Nima}  gives an explicit recursive formula for the all-loop integrand of scattering amplitudes.

The amplitudes in gauge theory are constrained by gauge symmetry. This leads to Ward identity which constrains the amplitudes at all loop level. Inspired by the BCFW momenta shift, we considered the Ward identity for tree level amplitudes with complexified momenta for a pair of external lines, and then obtained a recursion relation for the boundary terms using BCFW technique in our recent article \cite{Chen}. However, in \cite{Chen}, we chose a particular momenta shift such that the external states of the complexified lines are independent of the complex parameter $z$. Then a natural question is how to obtain a recursion relation for other possible momenta shifts. Furthermore, is it possible to obtain the full amplitudes from the Ward identity,  and to extend the technique to one loop amplitudes? In this article, we will give positive answers to all these questions. 

In section \ref{TreeRec}, we first give the proof of Ward identity at tree level using Feynman rules directly, and then derive the recursion relation for off shell amplitudes, where the cancellation details in the proof of Ward identity helps to simplify the recursion relations. Section \ref{LoopWard} is parallel to section \ref{TreeRec}. We first extend the proof of Ward identity to one loop level and then derive the recursion relation for one loop off shell amplitudes. Our technique does not rely on the on-shell momenta shifts. Also, in our calculation using the recursion relation, four point vertexes are not used explicitly. We calculate three and four point one loop off shell amplitudes as examples in section \ref{example}.

\section{Ward Identity and Implied Recursion Relation at Tree Level}\label{TreeRec}
In \cite{Chen}, we directly proved complexified Ward identity for pure Yang-Mills fields at tree level, and then used it to deduce a recursion relation for the boundary terms of the complexified amplitudes. Here we generalize the method to deduce a recursion relation for tree level amplitudes with one external off shell line. This section will serve as a basis for our generalization to one loop level in the next section. We will call the external off shell line $\Le$ with momentum $\ke^\mu$, and the corresponding off shell amplitudes $A_\mu$.

%We will also use the name "vector current" to denote $A_\mu$ in the text.

\subsection{Proof of Ward Identity at Tree Level}
Although done in our previous paper \cite{Chen}, we briefly summerize some key points in the direct proof of tree level Ward identity, since these points are useful for deriving tree level recursion relation and also will be part of the proof at one loop level in the next section.

The amplitude is complexified by shifting the momenta of a pair of external lines. We choose $\Le$ and one on shell line $L_s$ with momentum $k_s=\lmd_s\td\lmd_s$, and the shift is:
\beq
k_s \to k_s-z \eta,\ \ \ \ \ \ \ \ \ \ \ \ \ \ \ke\to \ke+z\eta,
\label{momshift}
\eeq
where z is the complexifing parameter and $\eta$ should satisfy $\eta^2=0$ and $k_s \cdot \eta=0$.

%\footnote{\textcolor{red}{should $\eta$ normal to the polarization vector?}}

The color ordered Feynman rules of the gauge field are as in \cite{Dixon1}, with outgoing momenta. We also write the Feynman rules for ghost fields here in Figure \ref{ghostFeynrule}, which will be used in the next section. 

\begin{figure}[htb]
\centering
\includegraphics[height=3.5cm,width=15cm]{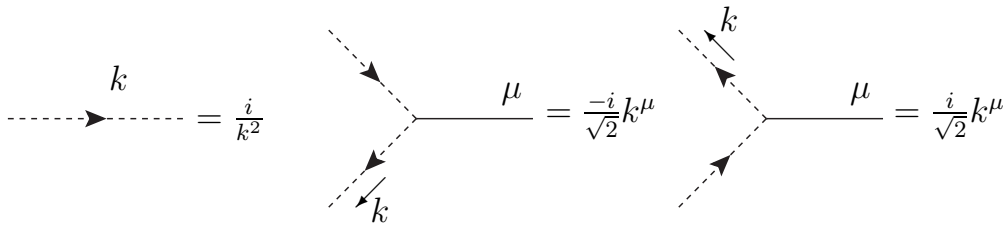}
\caption{Ghost field color ordered Feynman rules. Dashed line for ghost field and solid line for gauge field.}
\label{ghostFeynrule}
\end{figure}

For a three point vertex with line 1, 2 and $\Le$ in anti-clockwise order, we write it in the following form:
\bea\label{newV3}
V_{\mu_1\mu_2\mu}&\equiv&S_{\mu_1\mu_2\mu}+ R_{\mu_1\mu_2\mu}+M_{\mu_1\mu_2\mu},
\eea
where  
\bea\label{newV3s}
S_{\mu_1\mu_2\mu}&=&\frac{i}{\sqrt 2}\left(\eta_{\mu_1\mu_2}(k_1-k_2)_{\mu}\right) \nb\\
R_{\mu_1\mu_2\mu}&=&\frac{i}{\sqrt 2}\left(-2\eta_{\mu_2\mu}(\ke)_{\mu_1}+2\eta_{\mu\mu_1}(\ke)_{\mu_2}\right) \nb\\
M_{\mu_1\mu_2\mu}&=&\frac{i}{\sqrt 2}\left(-\eta_{\mu_2\mu}(k_1)_{\mu_1}+\eta_{\mu\mu_1}(k_2)_{\mu_2}\right). 
\eea
We will refer to these terms as S, R and M parts of the vertex. Contracting this vertex with $\ke$, we get:
\begin{equation}
\ke^\mu \cdot V_{\mu_1\mu_2\mu}=\frac{i}{\sqrt{2}}\eta_{\mu_1\mu_2} k_2^2-\frac{i}{\sqrt{2}}\eta_{\mu_1\mu_2} k_1^2+\frac{i}{\sqrt{2}}k_{2\ \mu_2}k_3{}_{\ \mu_1}-\frac{i}{\sqrt{2}}k_{1\ \mu_1}k_3{}_{\ \mu_2},
\label{kedotV}
\end{equation}
and we represent these terms by the symbols in Figure \ref{vertexnotation}. These terms are frequently used throughout the paper, and we will call the terms in the first line of Figure \ref{vertexnotation} as solid triangle terms, and the second line terms as hollow triangle terms.
\begin{figure}[htb]
\centering
\includegraphics[height=4cm,width=12cm]{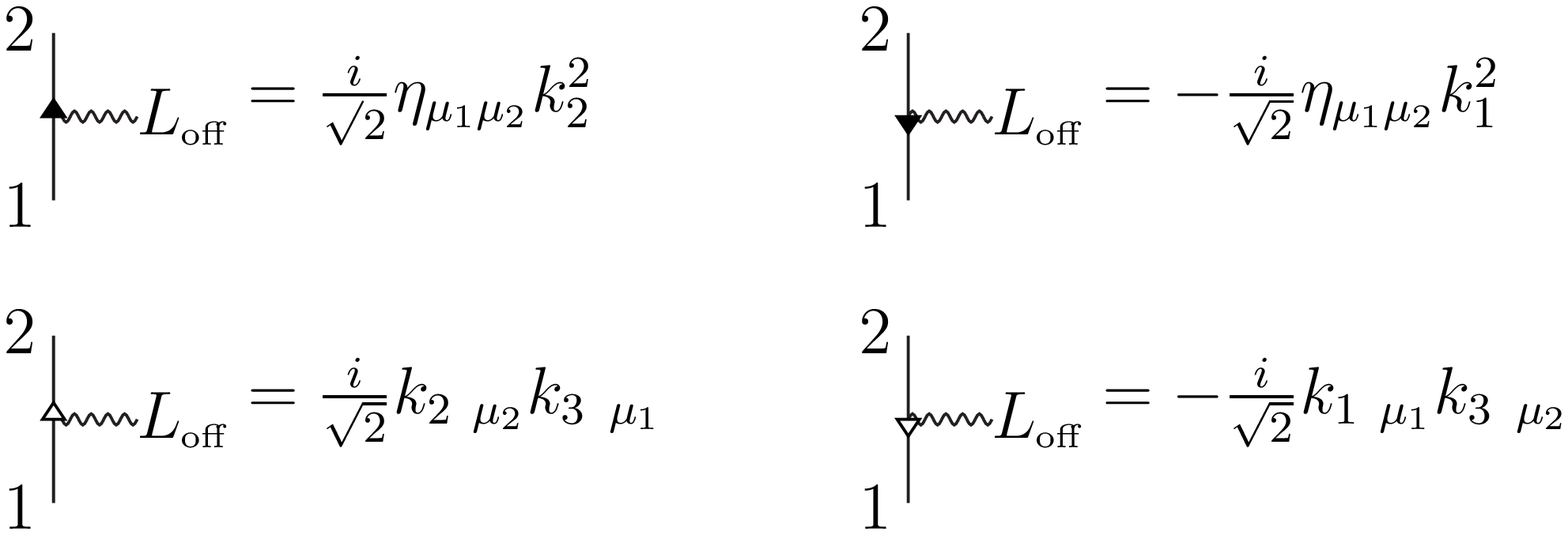}
\caption{Notations for \ref{kedotV}. We specialize $\Le$ using photon line.}
\label{vertexnotation}
\end{figure}

Then a proof of tree level Ward identity can be shown in several steps. Assume it holds for N-point and less than N-point amplitudes(for example, 3-point case can be immediately checked), we will show how it holds for (N+1) point amplitudes. We choose $\Le$ as the (N+1)-th line. We can construct an (N+1)-point color ordered diagram from an N-point one by inserting $\Le$ to an N-point diagram between Line 1 and Line N.

First, when $\Le$ is inserted to a propagator or Line 1 or Line N, we denote the vertex as $V_{\mbox{\tiny off}}$, and contract it with $\ke$, the following two hollow triangle terms in Figure \ref{treecancel1} vanish due to less-point Ward Identities or the on-shell conditions of Line 1 or N. The meaning of the symbols are in Figure \ref{vertexnotation}.

\begin{figure}[htb]
\centering
\includegraphics[height=2.5cm,width=10cm]{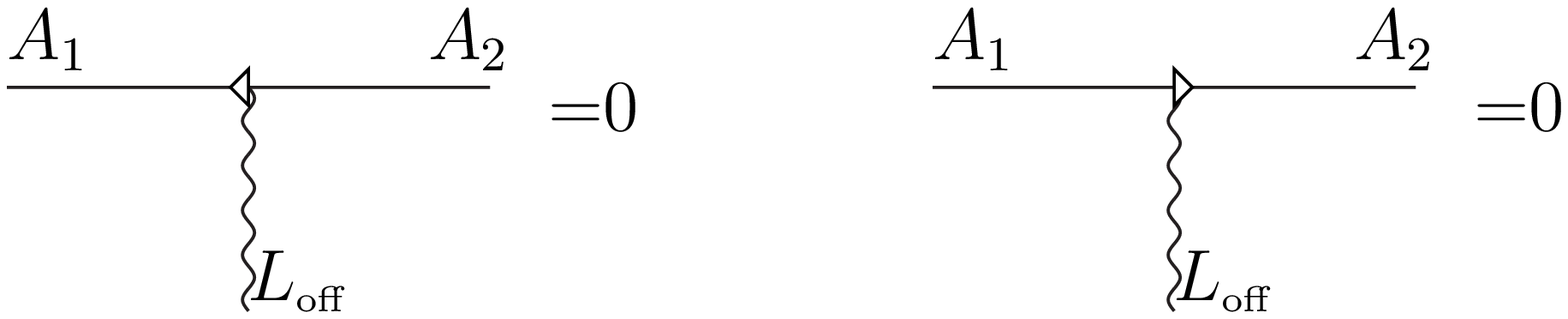}
\caption{When $\Le$ is inserted to a propagator or Line 1 or Line N, these terms vanish due to less point Ward identity or the on-shell conditions of Line 1 or N. $A_1$ and $A_2$ are sub amplitudes.}
\label{treecancel1}
\end{figure}

Second, $\Le$ is inserted to a three-point vertex in the N-point diagram. These terms and the remaining terms, ie. solid triangle terms, from the above case can be re-combined as in Figure \ref{treecancel2} to cancel each other.
\begin{figure}[htb]
\centering
\includegraphics[height=10cm,width=15cm]{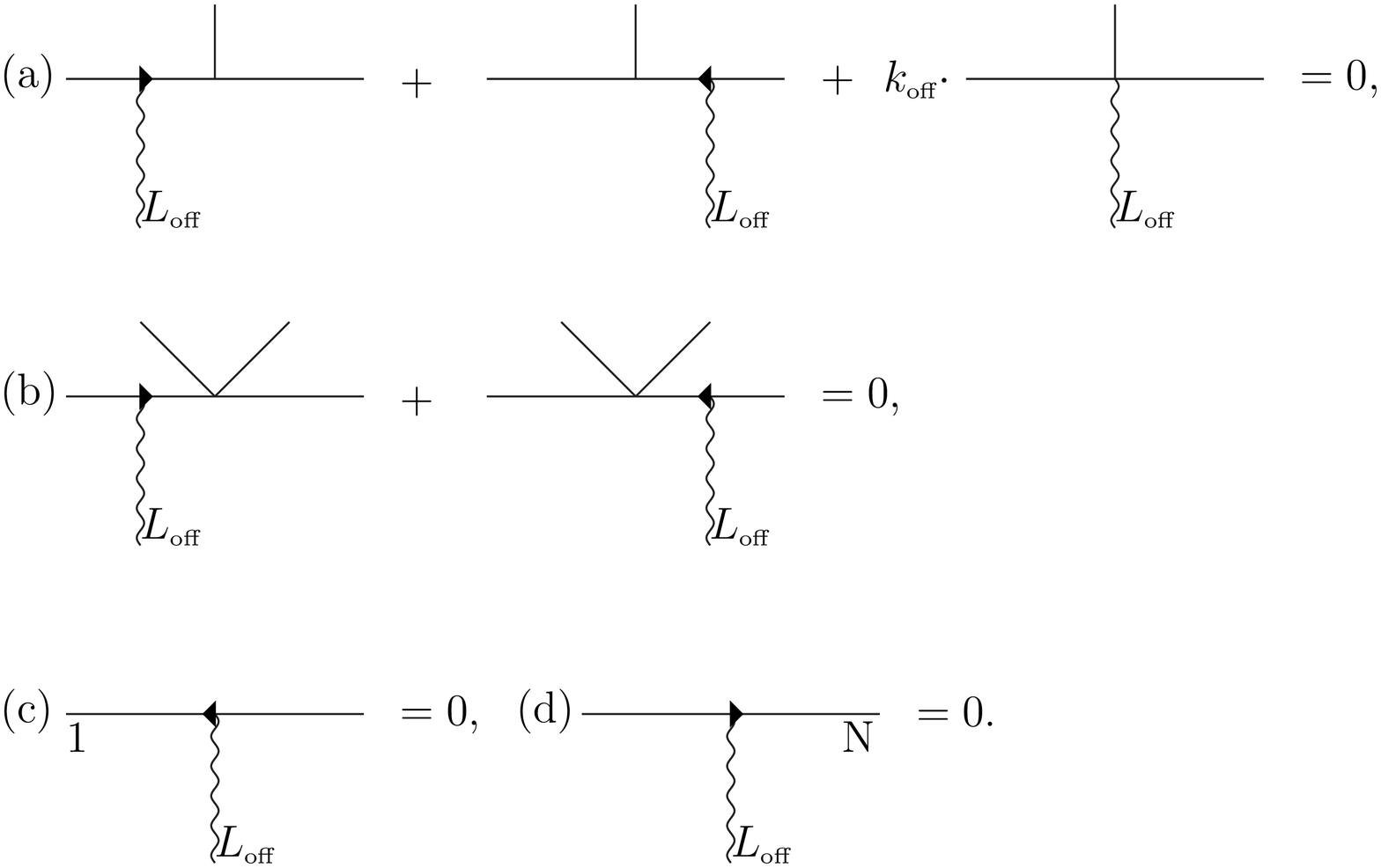}
\caption{A group of diagrams cancel. In (a) and (b), the cancellation is solely due to the vertex structures, not dependent on whether the legs are on shell or off shell. (c) and (d) are due to on shell conditions for Line 1 and Line N: $k_1^2=0$ and $k_N^2$=0.} 
\label{treecancel2}
\end{figure}

Figure \ref{treecancel1} and Figure \ref{treecancel2} constitute the proof of Ward identity at tree level.

\subsection{Recursion Relation for Tree Level Off Shell Amplitudes}\label{treerec}
As discussed in \cite{Chen}, from the complexified Ward identity ${{\hat k}_{\mbox{\tiny off}}}^\mu \cdot {\hat A}_\mu=0$, by a derivative over z we get:  
\be\label{recz}
\hat A_\mu \eta^\mu |_{z\to 0}=-{d\hat A_\mu\over dz}{\hat k}_{\mbox{\tiny off}}^\mu |_{z\to 0}.
\ee
The symbol $\hat{}$ represents that the quantity is complexified, ie. depends on the shift parameter z. Here $\ke$ is shifted as in \ref{momshift}: ${\hat k}_{\mbox{\tiny off}}=\ke+z\eta$. Our destination is to calculate $A_\mu$, and we will realize it by calculating the right hand side of \ref{recz}.

%For convenience we choose to shift $\Le$ and some other line $L_s$ which is not color adjacent to $\Le$. In other words, for a color ordered amplitude $A(1,2,\cdots,N,\Le)$, we choose to shift $\Le$ and another line which is not Line 1 or N. This is always possible for four and more point amplitudes, and for this kind of shift we need not consider the derivative of the polarization vector of $L_s$. The shift is as in \ref{momshift}.

We name the vertex which contains $\Le$ as $V_{\mbox{\tiny off}}$. At tree level, we have the following three cases: 
\begin{enumerate}
\item the derivative acts on a propagator;
\item the derivative acts on a three point vertex which does not contain $\Le$;
\item the derivative acts on a three point vertex $V_{\mbox{\tiny off}}$.
\end{enumerate}
In the first and second cases, when $V_{\mbox{\tiny off}}$ is a three point vertex, we write $\ke^\mu \cdot \Ve{}_{\ \mu}$ as in Figure \ref{vertexnotation}, and take out the hollow triangle terms. These terms, together with the terms from the third case where the derivative acts on the M part of $\Ve{}_{\ \mu}$ as written in \ref{newV3s}, add up to be 0 due to Ward identity for some sub amplitudes.

%add up to be proportional to some Ward Identities or the derivative of the Ward Identities of some sub-amplitudes, and thus add up to be 0.

From above we know that in the first and second cases, we only need the solid triangle terms for $\ke^\mu \cdot \Ve{}_{\ \mu}$ as represented in Figure \ref{vertexnotation}, when $\Ve$ is a three point vertex; in the third case, $\frac{d}{dz}$ only need to act on the S and R part of $\Ve{}_{\ \mu}$ as written in \ref{newV3s}. The first two cases can be further simplified. Due to (a) and (b) in Figure \ref{treecancel2}, the terms relevant for the first two cases are reduced to those with $\ke$ neighboring to the three point vertex or the propagator to be differentiated, as depicted in Figure \ref{Tree1}.

\begin{figure}[htb]
\centering
\includegraphics[height=13cm,width=10cm]{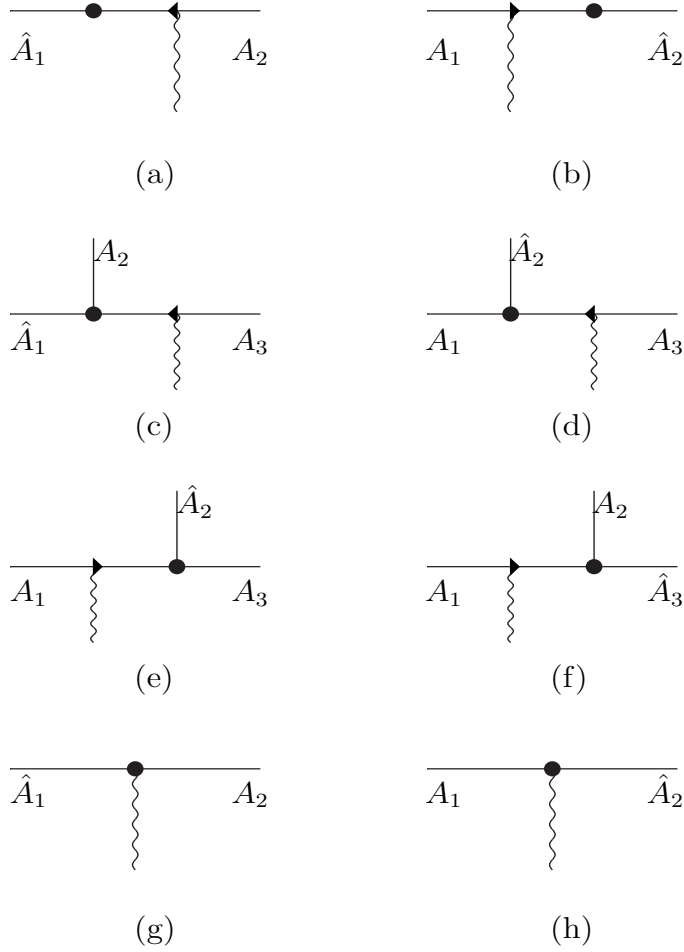}
\caption{Terms to be calculated for tree-level off-shell amplitudes. Here and following, the dark solid circle symbol $\bullet$ denotes where we act $d\over dz$. We shift $\Le$ and some other line $L_s$. $\{A_i\}$ denote some sub-diagrams with less external states. $\hat A_k$ includes $L_s$. In different diagrams, the same $A_k$ symbols do not mean the same sub amplitudes. They sum over all allowed sub amplitudes.}
\label{Tree1}
\end{figure}

%what about adjacent shift, is our formulas correct, and keeping $k_s\cdot \eta=0$ as if it is not 0 can or cannot give us the correct final result? It seems for adjacent shifting the diagrams and expressions are easier to write.

Thus, for the first case, the diagrams are (a) and (b) in Figure \ref{Tree1}. The contributions from (a) and (b) to $-{d\hat A_\mu\over dz}{\hat k}_{\mbox{\tiny off}}^\mu |_{z\to 0}$ are:
\bea
&&\mbox{for (a)}\ \ \frac{-i\sqrt{2}}{k_{A_1}^2 k_{A_2}^2}k_{A_1}\cdot \eta\  A_1\cdot A_2,\nonumber\\
&&\mbox{for (b)}\ \ \frac{i\sqrt{2}}{k_{A_1}^2 k_{A_2}^2}k_{A_2}\cdot \eta\  A_1\cdot A_2.
\label{preexptree1}
\eea
As noted in Figure \ref{Tree1}, \{$A_i$\} are some less point amplitudes. $k_{A_i}$ is the total momentum of the external legs contained in the sub amplitude $A_i$. If some $A_i$ just contains one external line $L_m$, we define this $A_i$ to be $i k_m^2 \epsilon_m$, and accompany it with a propagator $\frac{-i}{k_{A_i}^2}=\frac{-i}{k_m^2}$. Another point is that, although $k_s\cdot \eta=0$, we keep it in the evaluations here and below as if it is not 0, as will be explained at the end of this subsection.

The second case corresponds to (c) (d) (e) and (f) in Figure \ref{Tree1}, and the contributions to $-{d\hat A_\mu\over dz}{\hat k}_{\mbox{\tiny off}}^\mu |_{z\to 0}$ are:
\bea
&&\mbox{for (c)}\ \ \frac{1}{2 k_{A_1}^2 k_{A_2}^2 k_{A_3}^2}(A_3\cdot \eta\  A_1\cdot A_2+A_1\cdot \eta\  A_2\cdot A_3-2A_2\cdot \eta\  A_1\cdot A_3),\nonumber\\
&&\mbox{for (d)}\ \ \frac{1}{2 k_{A_1}^2 k_{A_2}^2 k_{A_3}^2}(-A_3\cdot \eta\  A_1\cdot A_2+2A_1\cdot \eta\  A_2\cdot A_3-A_2\cdot \eta\  A_1\cdot A_3),\nonumber\\
&&\mbox{for (e)}\ \ \frac{-1}{2 k_{A_1}^2 k_{A_2}^2 k_{A_3}^2}(-2A_3\cdot \eta\  A_1\cdot A_2+A_1\cdot \eta\  A_2\cdot A_3+A_2\cdot \eta\  A_1\cdot A_3),\nonumber\\
&&\mbox{for (f)}\ \ \frac{-1}{2 k_{A_1}^2 k_{A_2}^2 k_{A_3}^2}(-A_1\cdot \eta\  A_2\cdot A_3+2A_2\cdot \eta\  A_1\cdot A_3-A_3\cdot \eta\  A_1\cdot A_2).
\label{preexptree2}
\eea

And the third case corresponds to (g) and (h) in Figure \ref{Tree1}, whose contributions are:
\bea
&&\mbox{for (g)}\ \ \frac{-i}{\sqrt{2} k_{A_1}^2 k_{A_2}^2}(\ke\cdot \eta \ A_1\cdot A_2+2A_1\cdot \eta\  \ke\cdot A_2-2A_2\cdot \eta\  \ke\cdot A_1),\nonumber\\
&&\mbox{for (h)}\ \ \frac{-i}{\sqrt{2} k_{A_1}^2 k_{A_2}^2}(-\ke\cdot \eta\  A_1\cdot A_2+2A_1\cdot \eta\  \ke\cdot A_2-2A_2\cdot \eta\  \ke\cdot A_1).
\label{preexptree3}
\eea
As explained before \ref{preexptree1}, in this case $\frac{d}{dz}$ only need to act on the S and R part of $\Ve{}_{\ \mu}$ as written in \ref{newV3s}.

It can be observed that, \ref{preexptree2} for $L_s$ contained in $A_1$ or $A_2$ or $A_3$, the expressions are the same. In the case when $L_s$ is contained in $A_2$ we should sum (d) and (e) in Figure \ref{Tree1} to see that the expression is the same as when $L_s$ is contained in $A_1$ or $A_3$. The common expression is:
\beq
\frac{1}{2 k_{A_1}^2 k_{A_2}^2 k_{A_3}^2}(A_3\cdot \eta\  A_1\cdot A_2+A_1\cdot \eta\  A_2\cdot A_3-2A_2\cdot \eta\  A_1\cdot A_3).
\label{finaltree1}
\eeq
\ref{preexptree1} and \ref{preexptree2} summed up also give a common expression, regardless of whether $L_s$ is contained in $A_1$ or $A_2$:
\beq
\frac{-i}{\sqrt{2} k_{A_1}^2 k_{A_2}^2}\left(\ (k_{A_1}-k_{A_2})\cdot \eta\  A_1\cdot A_2+2A_1\cdot \eta\  \ke\cdot A_2-2A_2\cdot \eta\  \ke\cdot A_1\right)
\label{finaltree2}
\eeq
The final tree level result for $A_\mu \eta^\mu$ is the sum of \ref{finaltree1} and \ref{finaltree2}, which can be written in the form of $\tilde A_\mu \eta^\mu$. In the expressions we should sum over all allowed allocations of the on shell external legs into $\{A_i\}$. It is easy to show that the off shell amplitude $A_\mu=\tilde A_\mu$. In four dimensional spacetime, we only need to find 4 independent $\eta_i$ such that $A_\mu \eta^\mu_i=\tilde A_\mu \eta^\mu_i$. Since in the shift $\eta^\mu$ is required to satisfy $k_s \cdot \eta=0$ and $\eta^2=0$, the three choices of $\eta_i$ as $\epsilon_s^+$, $\epsilon_s^-$ or $k_s$ satisfy $A_\mu \eta^\mu_i=\tilde A_\mu \eta^\mu_i$. The remaining choice of $\eta_i$ can be chosen as $\ke$. This is not obvious to satisfy $A_\mu \eta^\mu_i=\tilde A_\mu \eta^\mu_i$. However, in our calculations we have kept the terms $k_s\cdot \eta$ as if it is not 0, and by this trick it comes out that $\tilde A_\mu \ke^\mu=0=A_\mu \ke^\mu$. In conclusion $A_\mu=\tilde A_\mu$, where $\tilde A_\mu$ is contained in the sum of \ref{finaltree1} and \ref{finaltree2} in form of $\tilde A_\mu \eta^\mu$.

Compare to Berends-Giele recursion relation \cite{Berends:1987me}, it is seen that \ref{finaltree1} corresponds to $\ke$ contained in a four point vertex, and \ref{finaltree2} is equivalent to the contribution when $\ke$ is contained in a three point vertex. This on one hand supports the correctness of our method, and on the other hand a little undermines the value of our method at tree level. There are also other recursion relations for off shell tree level amplitudes, eg. \cite{Feng}. Yet we are going to extend our method to one loop level, where the situation is much more complicated and our method is new. 

%proof read the above two paragraphs

%%%%%%%%%%%%%
\section{Ward Identity and Implied Recursion Relation at 1-loop Level}\label{LoopWard}

In this section we are going to extend our method to 1-loop level. We will show how complexified Ward identity holds at 1-loop level and then we deduce the corresponding recursive calculation of 1-loop off shell amplitude. Using our method, we will calculate three and four point 1-loop off shell amplitudes as examples. In our calculation we use FDH scheme \cite{Bern:1991aq}, in which only the loop momentum is continued to dimensionality different from 4.

We first explain some subtleties at loop level. First, after momentum shifting, some lines on the loop carry complex momenta. This brings ambiguities to the meaning of the loop integral and prevents us from translating the loop momentum $l\to l+k$ or flip it $l\to -l$. However, according to equation \ref{recz}, what we need for our technique is the derivative of the integral at the value $z\to 0$. And it is easy to prove that:
\begin{eqnarray}
&&\int d^D l \frac{d}{dz}f(l^\mu,{\hat k}^\mu)|_{z\to 0}=\int d^D l \frac{d}{dz}f(-l^\mu,{\hat k}^\mu)|_{z\to 0},\\
&&\int d^D l \frac{d}{dz}f(l^\mu,{\hat k}^\mu)|_{z\to 0}=\int d^D l \frac{d}{dz}f(l^\mu+{\hat k}'^\mu,{\hat k}^\mu)|_{z\to 0}.
\label{complexintegrand}
\end{eqnarray}
Thus for our technique, we can translate or flip the loop momentum even when the integrand is complex.

%for the fact that we only make the loop momentum in general dimension and external states in 4 dimension, what is the meaning of l+k, and in $l\to l+k$?

Second, some attention should be paid to color orderings and symmetry factors. At tree level there is only one color ordering contributing to the the primitive part of the color ordered amplitudes. At one loop level, most diagrams also only have one color ordering. However, for gauge field loop diagrams, there are three kinds of diagrams having two color orderings. Those are diagrams with two vertexes on the loop: two three-point vertexes; two four-point vertexes; a three-point vertex and a four-point vertex. For the first and second cases, the contributions from the two color orderings are the same at integrand level. For the third case, the contributions from the two color orderings at integrand level differ by a translation and flip of the loop momentum, and due to \ref{complexintegrand} the two orderings contribute the same in our method after integration. In a word, these three kinds of diagrams have a factor of 2 from possible color orderings. At the same time, these three kinds of diagrams have symmetry factor $\frac{1}{2}$, just canceling the doubling from color orderings. For ghost loop diagrams, those with two vertexes on the loop also have a doubling from two color orderings, while there is only either clockwise or anti clockwise ghost loop when there are only two vertexes on the ghost loop. We replace the doubling from color orderings by drawing both clockwise and anti-clockwise ghost loop diagrams, which are actually equal when there are only two vertexes on the ghost loop.

Finally, as our convention for the loop momentum for all our loop diagrams, we specify the loop momentum in the following way. For each external leg $L_i$, when we want to make a path from it to the loop, there is one definite vertex $V$ on the loop first encountered in the path, then we say the external leg $L_i$ is associated with this loop vertex $V$. We find the vertex with which $\Le$ is associated, call it $V_0$. Assume all the lines associated with $V_0$ in color ordering are $L_j, L_{j+1}, \cdots, L_N, \Le, L_1, \cdots, L_i$, then we assign the momentum of the first loop propagator on the counter clockwise side of $V_0$ as $l-k_1-\cdots-k_i$, with the loop momentum flowing in counter clockwise direction. $l$ is to be integrated. External leg momenta are outgoing.

\subsection{Proof of Ward Identity at 1-Loop Level}\label{loopWI}

In this section we use $A^l$ for 1-loop amplitudes, and $A^t$ for tree level amplitudes.

Two point and three point 1-loop Ward identity is easy to verify directly. Similar to the proof at tree level, we use induction, assume Ward identity holds for N and less than N point one loop amplitudes, and construct an (N+1) point diagram from an N point one by inserting $\ke$ in different places. We denote the vertex with $\ke$ as $V_{\mbox{\tiny off}}$ and when $V_{\mbox{\tiny off}}$ is a three point vertex, we decompose $\ke \cdot V_{\mbox{\tiny off}}$ as in Figure \ref{vertexnotation}.

{\bf Case 1.} When $\ke$ is linked to a propagator(including gauge field loop propagator) or external line of the N point diagram, the solid triangle terms from $\ke \cdot V_{\mbox{\tiny off}}$ mostly cancel the terms with $\ke$ in a four point vertex, in the manner of Figure \ref{treecancel2}. Only the terms in Figure \ref{loopremain1} remain. 
\begin{figure}[htb]
\centering
\includegraphics[height=3.8cm,width=13cm]{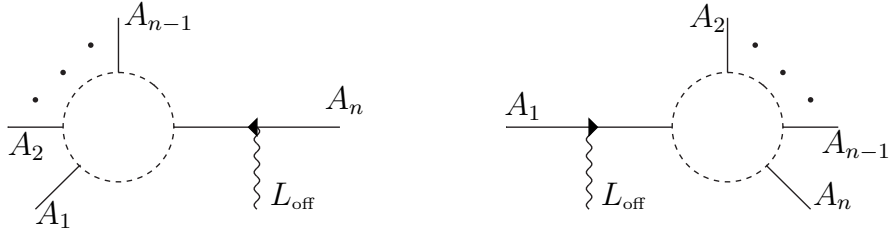}
\caption{The remaining terms in the first case that does not cancel in the manner of Figure \ref{treecancel2}. The loop is ghost loop and has two directions.}
\label{loopremain1}
\end{figure}

{\bf Case 2.} We need to consider the hollow triangle terms from $\ke \cdot V_{\mbox{\tiny off}}$ remaining from the above case, and we divide them into two sub cases:

\ \ {\bf \small Sub Case 1.} When $V_{\mbox{\tiny off}}$ is not on the loop, these terms vanish due to Ward identity for less point amplitudes in the induction assumption, similar to the tree level counterpart Figure \ref{treecancel1}.

\ \ {\bf \small Sub Case 2.} The remaining sub case is that $V_{\mbox{\tiny off}}$ is on the gauge field loop. We analyze one of the hollow triangle terms in Figure \ref{goontheloop}. The Figure has considered all the possible cases with the first right side vertex to be three or four point, and different types of second vertex relevant. When the first right side vertex is a three point vertex, acting on it with one of the factor in the hollow triangle term, we can again decompose it as in Figure \ref{vertexnotation} into solid and hollow triangle terms. (a) and (b) in Figure \ref{goontheloop} are in fact the same diagrams as in Figure \ref{treecancel2}. (c) vanishes due to tree level Ward identity, and (d) is due to on shell condition for external legs besides $\Le$. Then the type of term in (e) of Figure \ref{goontheloop} remains, which is a hollow triangle term staying on the loop, and it will act on the next vertex on the loop, repeating the same processes as in (a)-(d) of Figure \ref{goontheloop}, until it meets the final vertex on the loop. For this sub case, the remaining diagrams are in Figure \ref{loopremain2}.

\begin{figure}[htb]
\centering
\includegraphics[height=15cm,width=14cm]{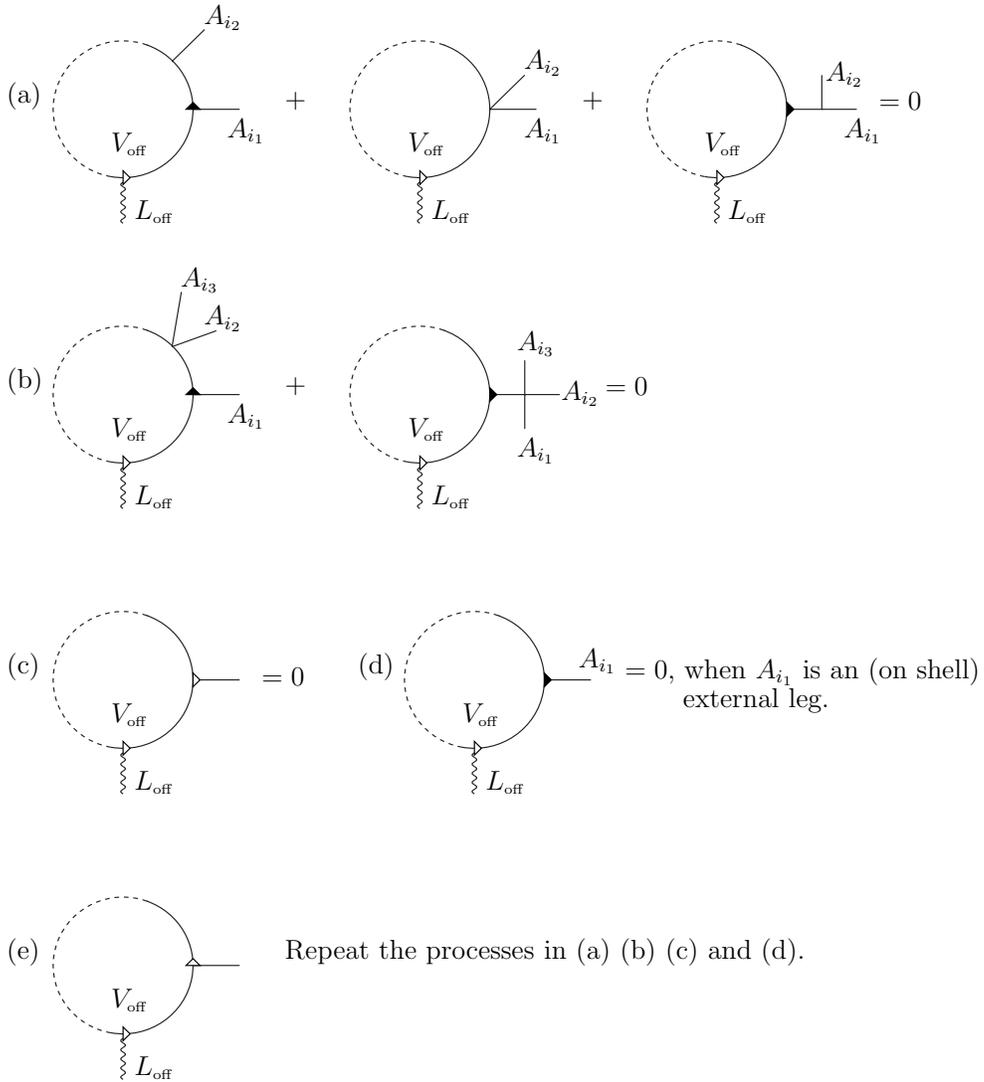}
\caption{Analysis of the action of the hollow triangle terms in {\bf\small Sub Case 2}. The dashed line is not ghost field, but just part of the loop diagram not relevant.}
\label{goontheloop}
\end{figure}

\begin{figure}[htb]
\centering
\includegraphics[height=8cm,width=12cm]{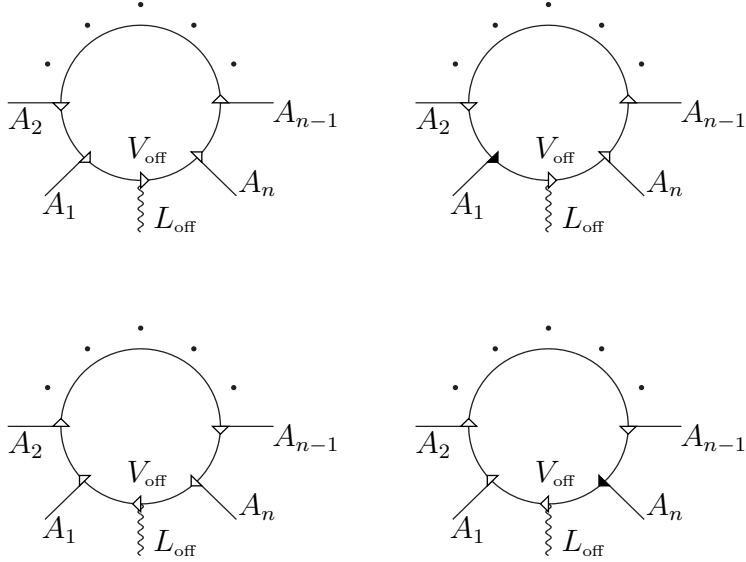}
\caption{The terms from {\bf \small Sub Case 2}. Except the hollow triangle terms at $V_{\mbox{\tiny off}}$, other hollow and solid triangle terms on the loop are induced from the hollow triangle term of the previous loop vertex, as described in the text of {\bf\small Sub Case 2.} and Figure \ref{goontheloop}.}
\label{loopremain2}
\end{figure}

%maybe we can use double triangle for kk term in k dot V expression. and then after it acts on the next vertex, we replace double triangle with single triangle.

{\bf Case 3.} The remaining case: $\ke$ is linked to a ghost propagator of the N point diagram, as in Figure \ref{loopremain3}.

\begin{figure}[htb]
\centering
\includegraphics[height=4cm,width=6cm]{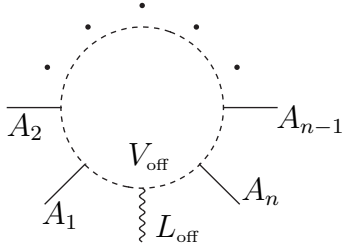}
\caption{Diagram for {\bf Case 3.} The ghost loop can be in two directions.}
\label{loopremain3}
\end{figure}

By direct and simple calculations, the terms from Figure \ref{loopremain1}, Figure \ref{loopremain2} and Figure \ref{loopremain3}, with same set of sub amplitudes $A_i$, add up to be 0. Combine {\bf Case 1, 2, 3}, we have proven that Ward identity holds at $N+1$ point one loop level. Thus by induction we have proven Ward identity holds at one loop level using Feynman rules in a direct way.

%The complication compared to tree level proof comes from the terms in Figure \ref{loopremain1}, Figure \ref{loopremain2} and Figure \ref{loopremain3}, where we find that the ghost loop contribution and a remaining part from the longitudinal component of the gluon vertex cancel.

\subsection{Recursion Relation for Loop Level Off Shell Amplitudes}\label{LoopRec}
Similar to the tree level off shell amplitudes calculation, we can use $\hat A_\mu \eta^\mu |_{z\to 0}=-{d\hat A_\mu\over dz}{\hat k}_{\mbox{\tiny off}}^\mu |_{z\to 0}$ to calculate one loop level off shell amplitudes. The experience at tree level, and the details of how Ward identity holds at one loop level discussed in the last subsection, help us to simplify our discussion and calculation of one loop level off shell amplitudes.

% Same as tree level, in order to avoid derivatives over the polarization vectors of external states, we choose to shift $\Le$ and one other line $L_s$ which is not color adjacent to $\Le$, which is always possible for four and more point amplitudes.

When the derivative acts on a gauge field propagator or a vertex which is not on the loop, we can use the expressions derived in section \ref{treerec} directly, ie. \ref{finaltree1} and \ref{finaltree2}, for the contribution to $-{d\hat A_\mu\over dz}{\hat k}_{\mbox{\tiny off}}^\mu |_{z\to 0}$:
\bea
&&\frac{1}{2 k_{A_1}^2 k_{A_2}^2 k_{A_3}^2}(A^{l/t}_3\cdot \eta\  A^{l/t}_1\cdot A^{l/t}_2+A^{l/t}_1\cdot \eta\  A^{l/t}_2\cdot A^{l/t}_3-2A^{l/t}_2\cdot \eta\  A^{l/t}_1\cdot A^{l/t}_3),\label{finalloop1}
\\
&&\frac{-i}{\sqrt{2} k_{A_1}^2 k_{A_2}^2}\left(\ (k_{A_1}-k_{A_2})\cdot \eta\  A^{l/t}_1\cdot A^{l/t}_2+2A^{l/t}_1\cdot \eta\  \ke\cdot A^{l/t}_2-2A^{l/t}_2\cdot \eta\  \ke\cdot A^{l/t}_1\right).\nonumber
\eea
In \ref{finalloop1}, we allocate the on shell external legs into $\{A^{l/t}_i\}$ in color ordering, with one and only one $A_i^{l/t}$ being one loop level. As in tree level, in each expression we should sum over all allowed allocations of the on shell external legs into $\{A^{l/t}_i\}$.

When the derivative acts on a gauge field loop propagator or a loop vertex, these are shown in Figure \ref{loop1}. For the same reasons as discussed in tree level recursion calculation, in (a) to (f), we only need to consider $\Le$ next to the propagator or vertex differentiated and only need the solid triangle term. In (g), we only differentiate the S and R terms of the vertex. The M part of the vertex in (g) will be dealt with in the following. In Figure \ref{loop1}, we encounter tree level two line off shell amplitudes $A^t_{\sigma\rho}$. This quantity can also be calculated recursively using our method, but in this paper we will not discuss it, and will use Feynman rules to calculate it in our example. Those $A^t$ without sub indices are tree level one line off shell amplitudes, which can apply our method in the previous section. (a) is 0 due to our convention for the loop momentum, described in the paragraph before section \ref{loopWI}.

Regardless of whether the other shifted line $L_s$ is among $\{L_1,L_2,\cdots,L_j\}$ or among $\{L_{j+1},\cdots,L_N\}$, the contributions to $-{d\hat A_\mu\over dz}{\hat k}_{\mbox{\tiny off}}^\mu |_{z\to 0}$ from Figure \ref{loop1} are (we use $K_{m,n}$ to represent for $k_m+k_{m+1}+\cdots+k_n$):
\bea
&&(a)\ \ \ \ \ \ \ \,:0\label{finalloop2}\\
&&(b)+(g): \frac{-i}{\sqrt{2} l^2 (l+\ke)^2}(\ (2 l+\ke)\cdot \eta\  A^t_{\sigma\rho}(1,2,\cdots,N) \ g^{\sigma\rho}\nonumber\\
&&\ \ \ \ \ \ \ \ \ \ \ \ \ +2\eta^\sigma\  A^t_{\sigma\rho}(1,2,\cdots,N) \ k^\rho-2\eta^\rho\  A^t_{\sigma\rho}(1,2,\cdots,N) \ k^\sigma)\nonumber\\
&&(d)+(e): \frac{1}{2 l^2(l-K_{1,j})^2 K_{j+1,N}^2}(A^t(j+1,\cdots,N)\cdot \eta\  A^t_{\sigma\rho}(1,2,\cdots,j) \ g^{\sigma\rho}\nonumber\\
&&\ \ \ \ \ \ \ \ \ \ \ \ \ -2A^{t\ \sigma}(j+1,\cdots,N) \ \eta^\rho \ A^t_{\sigma\rho}(1,2,\cdots,j) +A^{t\ \rho}(j+1,\cdots,N)\ \eta^\sigma\  A^t_{\sigma\rho}(1,2,\cdots,j)\ )\nonumber\\
&&(c)+(f):\frac{1}{2 (l+\ke)^2(l-K_{1,j})^2 K_{1,j}^2}(A^t(1,\cdots,j)\cdot \eta\  A^t_{\sigma\rho}(j+1,\cdots,N)\  g^{\sigma\rho}\nonumber\\
&&\ \ \ \ \ \ \ \ \ \ \ \ \ -2A^{t\ \rho}(1,\cdots,j) \ \eta^\sigma\  A^t_{\sigma\rho}(j+1,\cdots,N)  +A^{t\ \sigma}(1,\cdots,j)\ \eta^\rho\  A^t_{\sigma\rho}(j+1,\cdots,N)\ )\nonumber
\eea

\begin{figure}[htb]
\centering
\includegraphics[height=15cm,width=11cm]{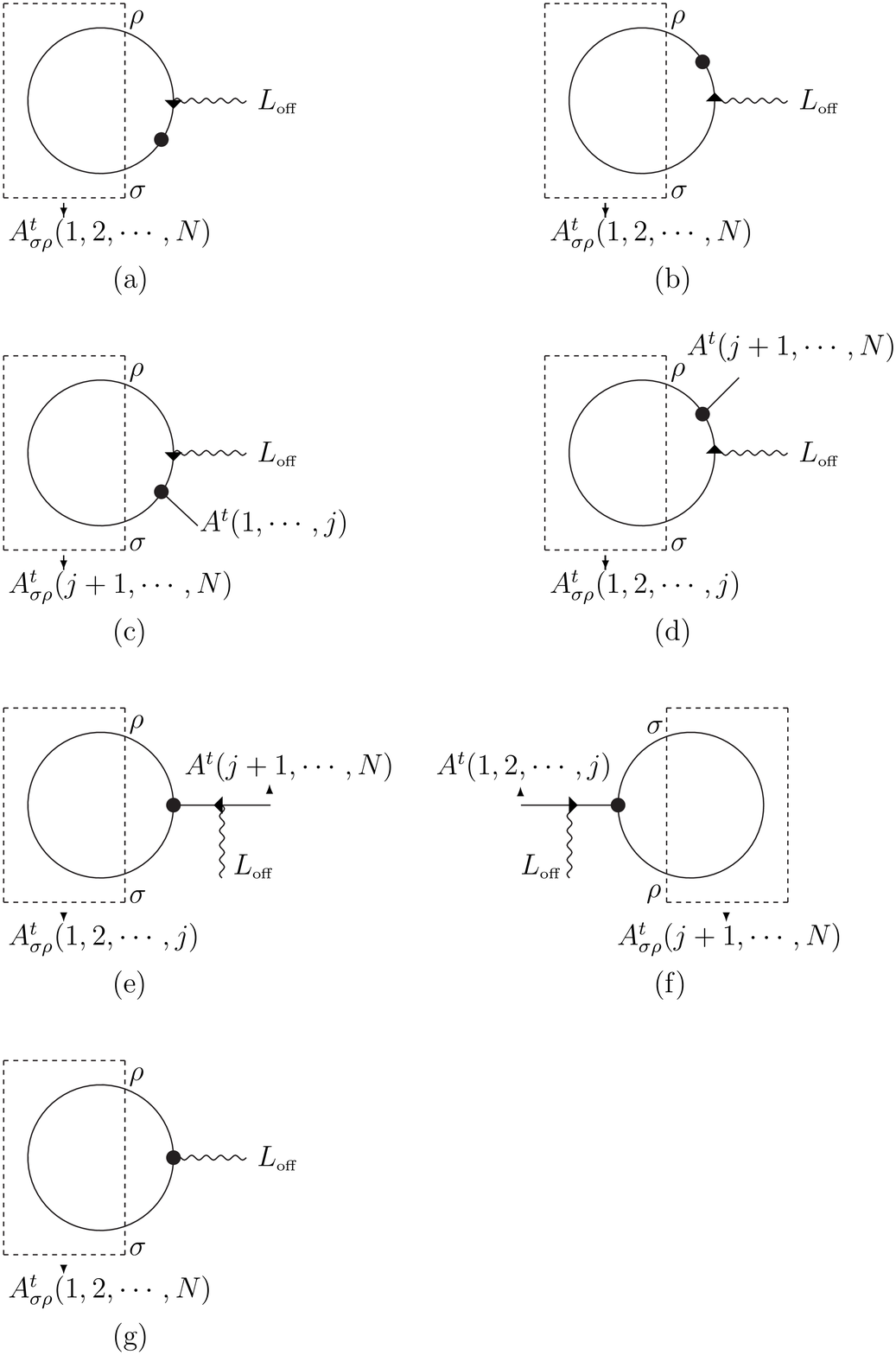}
\caption{Diagrams with derivative acting on the propagator or vertex on the loop, which cannot directly apply the tree level results.}
\label{loop1}
\end{figure}

The final contributions to $-{d\hat A^l_\mu\over dz} {{\hat k}_{\mbox{\tiny off}}}^\mu |_{z\to 0}$ come from the derivatives in the diagrams of Figure \ref{loopremain1}, Figure \ref{loopremain2} and Figure \ref{loopremain3}. Denoting the diagrams as $D_i$, since $\sum D_i\cdot \ke=0$ from the last subsection, we have $-\sum {d\hat D_{i\ \mu}\over dz}{\hat k}_{\mbox{\tiny off}}^\mu |_{z\to 0}=\sum  D_{i\ \mu}\  \eta^\mu$. This is like an opposite operation compared to the method in the current paper to deal with the set of diagrams $D_i$, but it simplifies the local calculation, eg. the diagrams in Figure \ref{loopremain1} turn out to be not contributing. We use $K_{A_{m,n}}$ to represent for $k_{A_m}+k_{A_{m+1}}+\cdots+k_{A_n}$, with $k_{A_i}$ the total momentum of the external legs contained in the sub amplitude $A_i$. The total momentum conservation is then $K_{A_{1,n}}+\ke=0$. The contributions to $-{d\hat A^l_\mu\over dz} {{\hat k}_{\mbox{\tiny off}}}^\mu |_{z\to 0}$ from derivatives in Figure \ref{loopremain1}, Figure \ref{loopremain2} and Figure \ref{loopremain3} are:
\bea
&&\frac{(-i)^n\ (l-K_{A_{1,1}})\cdot A_2\ (l-K_{A_{1,2}})\cdot A_3\ \cdots (l-K_{A_{1,n-2}})\cdot A_{n-1}}{(\sqrt{2})^{n+1}k_{A_1}^2\cdots k_{A_n}^2(l-K_{A_{1,1}})^2\cdots(l-K_{A_{1,n-1}})^2}\nonumber\\
&&(-\frac{2 l\cdot A_1\ (l-K_{A_{1,n-1}})\cdot A_n\ (2l+\ke)\cdot \eta}{l^2(l+\ke)^2}\nonumber\\
&&+\frac{l\cdot A_1\ A_n \cdot \eta}{l^2}+\frac{(l-K_{A_{1,n-1}})\cdot A_n \ A_1\cdot \eta}{(l+\ke)^2}).
\label{finalloop3}
\eea
This expression is well defined when $n\ge 2$. Especially when $n=2$, one should multiply the pre-factor with each term in the bracket to see that it is well defined. When $n=1$, the last two terms in the bracket vanish.

\ref{finalloop1}, \ref{finalloop2} and \ref{finalloop3} constitute our expressions for recursively calculating one loop off shell amplitudes. In each expression, eg. in the above one \ref{finalloop3}, we should sum over all the allowed different allocations of the on shell external legs into $A_1, \cdots, A_n$, with $n=1,2,\cdots,N$. This summation is not written explicitly in the expressions. Similar to our statement in the tree level counterpart, summing over \ref{finalloop1}, \ref{finalloop2} and \ref{finalloop3}, we get a form $A^l_\mu \eta^\mu$, with the $A^l_\mu$ our wanted one loop off shell amplitude.

\subsection{Examples of 1-loop Off Shell Amplitudes}\label{example}
As an application and verification of our method, we have computed three and four point one loop amplitudes with one off shell leg. ie. $A^l_\mu(k_1,k_2)$ and $A^l_\mu(k_1,k_2,k_3)$, by summing up the contributions from \ref{finalloop1}, \ref{finalloop2} and \ref{finalloop3}. We use the integral reduction method in \cite{Ellis} to reduce the integrals to scalar integrals. We use the following notations for the scalar integrations:
\beas
&&B0[1,3]=\int \frac{d^D l}{(2\pi)^D}\frac{1}{l^2(l-k_1-k_2)^2},\ \ B0[1,4]=\int \frac{d^D l}{(2\pi)^D}\frac{1}{l^2(l-k_1-k_2-k_3)^2},\\
&&B0[2,4]=\int \frac{d^D l}{(2\pi)^D}\frac{1}{l^2(l-k_2-k_3)^2},\ \ C0[1,2,3]=\int \frac{d^D l}{(2\pi)^D}\frac{1}{l^2(l-k_1)^2(l-k_1-k_2)^2},\\
&&C0[1,2,4]=\int \frac{d^D l}{(2\pi)^D}\frac{1}{l^2(l-k_1)^2(l-k_1-k_2-k_3)^2},\\
&&C0[1,3,4]=\int \frac{d^D l}{(2\pi)^D}\frac{1}{l^2(l-k_1-k_2)^2(l-k_1-k_2-k_3)^2},\\
&&C0[2,3,4]=\int \frac{d^D l}{(2\pi)^D}\frac{1}{l^2(l-k_2)^2(l-k_2-k_3)^2},\\
&&D0[1,2,3,4]=\int \frac{d^D l}{(2\pi)^D}\frac{1}{l^2(l-k_1)^2(l-k_1-k_2)^2(l-k_1-k_2-k_3)^2}.
\eeas
Other scalar integrations are not needed in this article. The evaluation of the scalar integrals see \cite{BernD2}.

We start from the two point function:
\begin{equation}
A^l_{\mu\nu}(k)=\frac{2-3D}{2(1-D)}(k^2 g_{\mu\nu}-k_\mu k_\nu) \int \frac{d^D l}{(2\pi)^D}\frac{1}{l^2(l-k)^2}.
\label{twopoint1loop}
\end{equation}

Then we can calculate three point one loop off shell amplitude using our method:
\bea
&&A^l_\mu(k_1,k_2)=\frac{1}{2\sqrt{2}}\left[k_1\cdot \epsilon_2 \ \epsilon_{1\ \mu}-k_2\cdot \epsilon_1\  \epsilon_{2\ \mu}-\frac{2D-5}{D-1}\epsilon_1\cdot \epsilon_2\  (k_1-k_2)_\mu\right.\nonumber\\
&&\ \ \ \ \ \ \ \ \ \ \ \ \ \ \ \ \ \ \ \ \ \ \ \ \ \left.+\frac{D-4}{D-1}\frac{k_1\cdot \epsilon_2\  k_2 \cdot \epsilon_1}{k_1\cdot k_2}(k_1-k_2)_\mu\right] B0[1,3]\label{threepoint1loop}\\
&&\ \ \ \ \ \ \ \ \ \ \ \ \ \ \ +\frac{1}{2\sqrt{2}}k_1\cdot k_2[-3 \epsilon_1\cdot \epsilon_2 \ (k_1-k_2)_\mu+4 k_1\cdot \epsilon_2 \ \epsilon_{1\ \mu}-4 k_2 \cdot \epsilon_1 \ \epsilon_{2\ \mu}]C0[1,2,3].\nonumber
\eea

At four point, the length of the expressions grow very quickly, and we will only give $A^l_\mu(1^+,2^+,3^+)$. Instead of giving this expression directly, we will give $A^l_\mu(1^+,2^+,3^+) \epsilon_1^\mu$, $A^l_\mu(1^+,2^+,3^+) \epsilon_3^\mu$ and $A^l_\mu(1^+,2^+,3^+) k_1^\mu$. Together with $A^l_\mu(1^+,2^+,3^+) (k_1+k_2+k_3)^\mu=0$, the expressions are enough to determine all the 4 components of $A^l_\mu(1^+,2^+,3^+)$. We choose the spinor representations for $k_{1,2,3}$ and $\epsilon_{1,2,3}$ to be:
\begin{equation}
k_1=\lambda_1 \tilde \lambda_1,\ k_2=\lambda_2 \tilde \lambda_2,\ k_3=\lambda_3 \tilde \lambda_3,\ \epsilon_1=\frac{\lambda_\nu \tilde \lambda_1}{\langle\lambda_\nu \lambda_1\rangle},\ \epsilon_2=\frac{\lambda_\nu \tilde \lambda_2}{\langle\lambda_\nu \lambda_2\rangle},\ \epsilon_3=\frac{\lambda_\nu \tilde \lambda_3}{\langle\lambda_\nu \lambda_3\rangle},
\end{equation}
with $\lambda_\nu$ an arbitrary reference spinor. We will use $\langle\nu 1\rangle$ to stand for $\langle\lambda_\nu \lambda_1\rangle$ and similarly others. We use $(A^l_\mu(1^+,2^+,3^+) \epsilon_1^\mu)_{D0[1,2,3,4]}$ to denote for the coefficient of $D0[1,2,3,4]$ in $A^l_\mu(1^+,2^+,3^+) \epsilon_1^\mu$, and similarly for others. We give the coefficients at $D=4$. The off shell line makes the expressions much more complicated than that with all on shell lines. On one hand, when all lines are on shell, since the amplitudes are gauge invariant, we can choose some specific reference spinor, while in the off shell case we should keep the reference spinor $\lambda_\nu$ arbitrary. On the other hand, there are many terms in the expressions below which is 0 when all lines are on shell. For example, the first coefficient below would be 0 due to $(\langle 12 \rangle [12]+\langle 13 \rangle [13]+\langle 23 \rangle [23])=0$ when all lines were on shell.

%To simplify the expressions, we use spinors and vectors in a mixed way. $k_4=-k_1-k_2-k_3$.

Then for $A^l_\mu(1^+,2^+,3^+) \epsilon_1^\mu$:
\begin{scriptsize}
\beas
&&(A^l_\mu(1^+,2^+,3^+) \epsilon_1^\mu)_{D0[1,2,3,4]}\\
&&=-\frac{(\langle 13 \rangle \langle 2\nu \rangle-2 \langle 12 \rangle \langle 3\nu \rangle) [12] (\langle 2\nu \rangle [12]+\langle 3\nu \rangle [13]) [23] (\langle 12 \rangle [12]+\langle 13 \rangle [13]+\langle 23 \rangle [23])}{64 \langle 13 \rangle^2 \langle 1\nu \rangle \langle 2\nu \rangle},\\
&&\\
&&(A^l_\mu(1^+,2^+,3^+) \epsilon_1^\mu)_{C0[1,2,4]}\\
&&=-\frac{(\langle 2\nu \rangle [12]+\langle 3\nu \rangle [13])}{64 \langle 12 \rangle \langle 13 \rangle^2 \langle 1\nu \rangle^2 \langle 23 \rangle \langle 2\nu \rangle \langle 3\nu \rangle (\langle 12 \rangle [12]+\langle 13 \rangle [13])} (\langle 1\nu \rangle^2 \langle 23 \rangle^2 (\langle 12 \rangle \langle 13 \rangle \langle 2\nu \rangle [12]^2+2 \langle 12 \rangle \langle 12 \rangle \langle 3\nu \rangle [12]^2\\
&&+4 \langle 12 \rangle \langle 13 \rangle \langle 3\nu \rangle [13] [12]+\langle 13 \rangle^2 \langle 3\nu \rangle [13]^2) [23]-\langle 1\nu \rangle (\langle 12 \rangle [12]+\langle 13 \rangle [13]) (2 \langle 3\nu \rangle^2 [12]^2 \langle 12 \rangle^3+\langle 13 \rangle \langle 3\nu \rangle [12] (4 \langle 3\nu \rangle [13]\\
&&-\langle 2\nu \rangle [12]) \langle 12 \rangle^2-2 \langle 13 \rangle^2 \langle 2\nu \rangle [12] (\langle 2\nu \rangle [12]+3 \langle 3\nu \rangle [13]) \langle 12 \rangle-\langle 13 \rangle^3 \langle 2\nu \rangle \langle 3\nu \rangle [13]^2)),\\
&&\\
&&(A^l_\mu(1^+,2^+,3^+) \epsilon_1^\mu)_{C0[2,3,4]}\\
&&=-\frac{(\langle 2\nu \rangle [12]+\langle 3\nu \rangle [13]) [23] }{64 \langle 12 \rangle \langle 13 \rangle^2 \langle 1\nu \rangle^2 \langle 2\nu \rangle}(2 \langle 3\nu \rangle (\langle 1\nu \rangle [12]-\langle 3\nu \rangle [23]) \langle 12 \rangle^2+\langle 13 \rangle (-\langle 1\nu \rangle \langle 2\nu \rangle [12]+2 \langle 1\nu \rangle \langle 3\nu \rangle [13]\\
&&+3 \langle 2\nu \rangle \langle 3\nu \rangle [23]) \langle 12 \rangle+\langle 13 \rangle^2 \langle 2\nu \rangle (\langle 1\nu \rangle [13]+\langle 2\nu \rangle [23])),\\
&&\\
&&(A^l_\mu(1^+,2^+,3^+) \epsilon_1^\mu)_{C0[1,3,4]}\\
&&=\frac{(\langle 2\nu \rangle [12]+\langle 3\nu \rangle [13])}{64 \langle 12 \rangle \langle 13 \rangle^2 \langle 1\nu \rangle^2 \langle 23 \rangle \langle 2\nu \rangle (\langle 13 \rangle [13]+\langle 23 \rangle [23])} (-2 \langle 3\nu \rangle^2 [12] [23] (2 \langle 13 \rangle [13]+\langle 23 \rangle [23]) \langle 12 \rangle^3+\langle 3\nu \rangle (-2 \langle 23 \rangle^2 \langle 3\nu \rangle [23]^3\\
&&+3 \langle 13 \rangle \langle 23 \rangle (\langle 2\nu \rangle [12]-2 \langle 3\nu \rangle [13]) [23]^2+\langle 13 \rangle^2 [13] (\langle 1\nu \rangle [12] [13]-4 \langle 3\nu \rangle [23] [13]+4 \langle 2\nu \rangle [12] [23])) \langle 12 \rangle^2+\langle 13 \rangle \langle 2\nu \rangle \langle 3\nu \rangle [23] (\langle 13 \rangle^2 [13]^2\\
&&+5 \langle 13 \rangle \langle 23 \rangle [23] [13]+3 \langle 23 \rangle^2 [23]^2) \langle 12 \rangle+\langle 13 \rangle^2 \langle 2\nu \rangle (\langle 23 \rangle^2 \langle 2\nu \rangle [23]^3+3 \langle 13 \rangle \langle 23 \rangle \langle 2\nu \rangle [13] [23]^2+\langle 13 \rangle^2 [13]^2 (\langle 1\nu \rangle [13]+3 \langle 2\nu \rangle [23]))),\\
&&\\
&&(A^l_\mu(1^+,2^+,3^+) \epsilon_1^\mu)_{C0[1,2,3]}\\
&&=\frac{[12] (\langle 2\nu \rangle [12]+\langle 3\nu \rangle [13])}{64 \langle 13 \rangle^2 \langle 1\nu \rangle^2 \langle 23 \rangle \langle 2\nu \rangle \langle 3\nu \rangle} (\langle 2\nu \rangle (2 \langle 1\nu \rangle \langle 2\nu \rangle [12]+3 \langle 1\nu \rangle \langle 3\nu \rangle [13]+\langle 2\nu \rangle \langle 3\nu \rangle [23]) \langle 13 \rangle^2\\
&&+\langle 12 \rangle \langle 3\nu \rangle (\langle 1\nu \rangle \langle 2\nu \rangle [12]-2 \langle 1\nu \rangle \langle 3\nu \rangle [13]-3 \langle 2\nu \rangle \langle 3\nu \rangle [23]) \langle 13 \rangle+2 \langle 12 \rangle^2 \langle 3\nu \rangle^2 (\langle 3\nu \rangle [23]-\langle 1\nu \rangle [12])),\\
&&\\
&&(A^l_\mu(1^+,2^+,3^+) \epsilon_1^\mu)_{B0[2,4]}\\
&&=-\frac{(\langle 2\nu \rangle [12]+\langle 3\nu \rangle [13])}{64 \langle 12 \rangle \langle 13 \rangle \langle 1\nu \rangle^2 \langle 23 \rangle \langle 2\nu \rangle \langle 3\nu \rangle (\langle 12 \rangle [12]+\langle 13 \rangle [13])^2} (4 \langle 3\nu \rangle^2 [12] [13] (\langle 1\nu \rangle [12]-\langle 3\nu \rangle [23]) \langle 12 \rangle^3+\langle 13 \rangle (\langle 1\nu \rangle [12] (2 \langle 2\nu \rangle^2 [12]^2\\
&&-6 \langle 2\nu \rangle \langle 3\nu \rangle [13] [12]+5 \langle 3\nu \rangle^2 [13]^2)+\langle 3\nu \rangle (-4 \langle 2\nu \rangle^2 [12]^2+7 \langle 2\nu \rangle \langle 3\nu \rangle [13] [12]-2 \langle 3\nu \rangle^2 [13]^2) [23]) \langle 12 \rangle^2\\
&&+\langle 13 \rangle^2 (\langle 1\nu \rangle [13] (7 \langle 2\nu \rangle^2 [12]^2-6 \langle 2\nu \rangle \langle 3\nu \rangle [13] [12]+\langle 3\nu \rangle^2 [13]^2)+\langle 2\nu \rangle (3 \langle 2\nu \rangle^2 [12]^2-10 \langle 2\nu \rangle \langle 3\nu \rangle [13] [12]\\
&&+\langle 3\nu \rangle^2 [13]^2) [23]) \langle 12 \rangle+5 \langle 13 \rangle^3 \langle 2\nu \rangle^2 [12] [13] (\langle 1\nu \rangle [13]+\langle 2\nu \rangle [23])),\\
&&\\
&&(A^l_\mu(1^+,2^+,3^+) \epsilon_1^\mu)_{B0[1,4]}\\
&&=\frac{(\langle 2\nu \rangle [12]+\langle 3\nu \rangle [13])}{64 \langle 12 \rangle \langle 13 \rangle \langle 1\nu \rangle^2 \langle 23 \rangle \langle 2\nu \rangle \langle 3\nu \rangle (\langle 12 \rangle [12]+\langle 13 \rangle [13])^2 (\langle 13 \rangle [13]+\langle 23 \rangle [23])^3} (-2 \langle 3\nu \rangle^3 [12]^3 [13] [23] (3 \langle 13 \rangle [13]+2 \langle 23 \rangle [23]) \langle 12 \rangle^5\\
&&+\langle 3\nu \rangle^2 [12]^2 [13] (-8 \langle 23 \rangle^2 \langle 3\nu \rangle [23]^3+\langle 13 \rangle \langle 23 \rangle (5 \langle 2\nu \rangle [12]-28 \langle 3\nu \rangle [13]) [23]^2+2 \langle 13 \rangle^2 [13] (\langle 1\nu \rangle [12] [13]-14 \langle 3\nu \rangle [23] [13]\\
&&+5 \langle 2\nu \rangle [12] [23])) \langle 12 \rangle^4-\langle 3\nu \rangle [12] (4 \langle 23 \rangle^3 \langle 3\nu \rangle^2 [13] [23]^4+4 \langle 13 \rangle \langle 23 \rangle^2 (\langle 2\nu \rangle^2 [12]^2-3 \langle 2\nu \rangle \langle 3\nu \rangle [13] [12]+6 \langle 3\nu \rangle^2 [13]^2) [23]^3\\
&&+2 \langle 13 \rangle^2 \langle 23 \rangle [13] (6 \langle 2\nu \rangle [12]-11 \langle 3\nu \rangle [13]) (\langle 2\nu \rangle [12]-2 \langle 3\nu \rangle [13]) [23]^2+\langle 13 \rangle^3 [13]^2 (\langle 1\nu \rangle [12] [13] (3 \langle 2\nu \rangle [12]-8 \langle 3\nu \rangle [13])\\
&&+2 (6 \langle 2\nu \rangle^2 [12]^2-28 \langle 2\nu \rangle \langle 3\nu \rangle [13] [12]+15 \langle 3\nu \rangle^2 [13]^2) [23])) \langle 12 \rangle^3+\langle 13 \rangle (\langle 23 \rangle^3 \langle 3\nu \rangle (-4 \langle 2\nu \rangle^2 [12]^2+7 \langle 2\nu \rangle \langle 3\nu \rangle [13] [12]\\
&&-2 \langle 3\nu \rangle^2 [13]^2) [23]^4+\langle 13 \rangle \langle 23 \rangle^2 (\langle 2\nu \rangle^3 [12]^3-32 \langle 2\nu \rangle^2 \langle 3\nu \rangle [13] [12]^2+33 \langle 2\nu \rangle \langle 3\nu \rangle^2 [13]^2 [12]-10 \langle 3\nu \rangle^3 [13]^3) [23]^3\\
&&+\langle 13 \rangle^2 \langle 23 \rangle [13] (3 \langle 2\nu \rangle^3 [12]^3-72 \langle 2\nu \rangle^2 \langle 3\nu \rangle [13] [12]^2+62 \langle 2\nu \rangle \langle 3\nu \rangle^2 [13]^2 [12]-14 \langle 3\nu \rangle^3 [13]^3) [23]^2+\langle 13 \rangle^3 [13]^2 (\langle 1\nu \rangle [12] [13] (\langle 2\nu \rangle^2 [12]^2\\
&&-14 \langle 2\nu \rangle \langle 3\nu \rangle [13] [12]+6 \langle 3\nu \rangle^2 [13]^2)+(3 \langle 2\nu \rangle^3 [12]^3-64 \langle 2\nu \rangle^2 \langle 3\nu \rangle [13] [12]^2+53 \langle 2\nu \rangle \langle 3\nu \rangle^2 [13]^2 [12]-6 \langle 3\nu \rangle^3 [13]^3) [23])) \langle 12 \rangle^2\\
&&+\langle 13 \rangle^2 \langle 2\nu \rangle (\langle 23 \rangle^3 (3 \langle 2\nu \rangle^2 [12]^2-10 \langle 2\nu \rangle \langle 3\nu \rangle [13] [12]+\langle 3\nu \rangle^2 [13]^2) [23]^4+\langle 13 \rangle \langle 23 \rangle^2 [13] (\langle 3\nu \rangle [13]-15 \langle 2\nu \rangle [12]) (3 \langle 3\nu \rangle [13]\\
&&-\langle 2\nu \rangle [12]) [23]^3+3 \langle 13 \rangle^2 \langle 23 \rangle [13]^2 (9 \langle 2\nu \rangle^2 [12]^2-26 \langle 2\nu \rangle \langle 3\nu \rangle [13] [12]+\langle 3\nu \rangle^2 [13]^2) [23]^2+\langle 13 \rangle^3 [13]^3 (\langle 1\nu \rangle [12] [13] (6 \langle 2\nu \rangle [12]\\
&&-11 \langle 3\nu \rangle [13])+(21 \langle 2\nu \rangle^2 [12]^2-58 \langle 2\nu \rangle \langle 3\nu \rangle [13] [12]+\langle 3\nu \rangle^2 [13]^2) [23])) \langle 12 \rangle+5 \langle 13 \rangle^3 \langle 2\nu \rangle^2 [12] [13] (\langle 23 \rangle^3 \langle 2\nu \rangle [23]^4\\
&&+4 \langle 13 \rangle \langle 23 \rangle^2 \langle 2\nu \rangle [13] [23]^3+6 \langle 13 \rangle^2 \langle 23 \rangle \langle 2\nu \rangle [13]^2 [23]^2+\langle 13 \rangle^3 [13]^3 (\langle 1\nu \rangle [13]+4 \langle 2\nu \rangle [23]))),\\
&&\\
&&(A^l_\mu(1^+,2^+,3^+) \epsilon_1^\mu)_{B0[1,3]}\\
&&=-\frac{(\langle 2\nu \rangle [12]+\langle 3\nu \rangle [13])}{64 \langle 12 \rangle \langle 13 \rangle \langle 1\nu \rangle^2 \langle 23 \rangle \langle 2\nu \rangle \langle 3\nu \rangle (\langle 13 \rangle [13]+\langle 23 \rangle [23])^3} (\langle 2\nu \rangle [13]^2 (\langle 1\nu \rangle [13] (\langle 2\nu \rangle [12]-\langle 3\nu \rangle [13])+\langle 2\nu \rangle (3 \langle 2\nu \rangle [12]\\
&&-5 \langle 3\nu \rangle [13]) [23]) \langle 13 \rangle^4+[13] (3 \langle 23 \rangle \langle 2\nu \rangle^2 (\langle 2\nu \rangle [12]-3 \langle 3\nu \rangle [13]) [23]^2+\langle 12 \rangle \langle 3\nu \rangle [13] (\langle 1\nu \rangle [13] (\langle 3\nu \rangle [13]-3 \langle 2\nu \rangle [12])\\
&&+6 \langle 2\nu \rangle (\langle 3\nu \rangle [13]-2 \langle 2\nu \rangle [12]) [23])) \langle 13 \rangle^3+(\langle 23 \rangle^2 \langle 2\nu \rangle^2 (\langle 2\nu \rangle [12]-7 \langle 3\nu \rangle [13]) [23]^3+6 \langle 12 \rangle \langle 23 \rangle \langle 2\nu \rangle \langle 3\nu \rangle [13] (\langle 3\nu \rangle [13]\\
&&-2 \langle 2\nu \rangle [12]) [23]^2+2 \langle 12 \rangle^2 \langle 3\nu \rangle^2 [13]^2 (\langle 1\nu \rangle [12] [13]-3 \langle 3\nu \rangle [23] [13]+5 \langle 2\nu \rangle [12] [23])) \langle 13 \rangle^2-\langle 3\nu \rangle [23] (6 \langle 3\nu \rangle^2 [12] [13]^2 \langle 12 \rangle^3\\
&&+\langle 23 \rangle \langle 3\nu \rangle [13] (9 \langle 3\nu \rangle [13]-5 \langle 2\nu \rangle [12]) [23] \langle 12 \rangle^2+2 \langle 23 \rangle^2 \langle 2\nu \rangle (2 \langle 2\nu \rangle [12]-\langle 3\nu \rangle [13]) [23]^2 \langle 12 \rangle+2 \langle 23 \rangle^3 \langle 2\nu \rangle^2 [23]^3) \langle 13 \rangle\\
&&-4 \langle 12 \rangle^2 \langle 23 \rangle \langle 3\nu \rangle^3 [13] [23]^2 (\langle 12 \rangle [12]+\langle 23 \rangle [23])).
\eeas
\end{scriptsize}

For $A^l_\mu(1^+,2^+,3^+) \epsilon_3^\mu$:
\begin{scriptsize}
\beas
&&(A^l_\mu(1^+,2^+,3^+) \epsilon_3^\mu)_{D0[1,2,3,4]}\\
&&=\frac{(\langle 13 \rangle \langle 2\nu \rangle-2 \langle 12 \rangle \langle 3\nu \rangle) [12] [23] (\langle 1\nu \rangle [13]+\langle 2\nu \rangle [23]) (\langle 12 \rangle [12]+\langle 13 \rangle [13]+\langle 23 \rangle [23])}{64 \langle 13 \rangle^2 \langle 2\nu \rangle \langle 3\nu \rangle},\\
&&\\
&&(A^l_\mu(1^+,2^+,3^+) \epsilon_3^\mu)_{C0[1,2,4]}\\
&&=\frac{(\langle 1\nu \rangle [13]+\langle 2\nu \rangle [23])}{64 \langle 12 \rangle \langle 13 \rangle^2 \langle 1\nu \rangle \langle 23 \rangle \langle 2\nu \rangle \langle 3\nu \rangle^2 (\langle 12 \rangle [12]+\langle 13 \rangle [13])} (\langle 1\nu \rangle^2 \langle 23 \rangle^2 (\langle 12 \rangle (\langle 13 \rangle \langle 2\nu \rangle+2 \langle 12 \rangle \langle 3\nu \rangle) [12]^2+4 \langle 12 \rangle \langle 13 \rangle \langle 3\nu \rangle [13] [12]\\
&&+\langle 13 \rangle^2 \langle 3\nu \rangle [13]^2) [23]-\langle 1\nu \rangle (\langle 12 \rangle [12]+\langle 13 \rangle [13]) (2 \langle 3\nu \rangle^2 [12]^2 \langle 12 \rangle^3+\langle 13 \rangle \langle 3\nu \rangle [12] (4 \langle 3\nu \rangle [13]-\langle 2\nu \rangle [12]) \langle 12 \rangle^2\\
&&-2 \langle 13 \rangle^2 \langle 2\nu \rangle [12] (\langle 2\nu \rangle [12]+3 \langle 3\nu \rangle [13]) \langle 12 \rangle-\langle 13 \rangle^3 \langle 2\nu \rangle \langle 3\nu \rangle [13]^2)),\\
&&\\
&&(A^l_\mu(1^+,2^+,3^+) \epsilon_3^\mu)_{C0[2,3,4]}\\
&&=\frac{[23] (\langle 1\nu \rangle [13]+\langle 2\nu \rangle [23])}{64 \langle 12 \rangle \langle 13 \rangle^2 \langle 1\nu \rangle \langle 2\nu \rangle \langle 3\nu \rangle} (2 \langle 3\nu \rangle (\langle 1\nu \rangle [12]-\langle 3\nu \rangle [23]) \langle 12 \rangle^2+\langle 13 \rangle (-\langle 1\nu \rangle \langle 2\nu \rangle [12]+2 \langle 1\nu \rangle \langle 3\nu \rangle [13]+3 \langle 2\nu \rangle \langle 3\nu \rangle [23]) \langle 12 \rangle\\
&&+\langle 13 \rangle^2 \langle 2\nu \rangle (\langle 1\nu \rangle [13]+\langle 2\nu \rangle [23])) ,\\
&&\\
&&(A^l_\mu(1^+,2^+,3^+) \epsilon_3^\mu)_{C0[1,3,4]}\\
&&=-\frac{1}{64 \langle 12 \rangle \langle 13 \rangle^2 \langle 1\nu \rangle \langle 23 \rangle \langle 2\nu \rangle \langle 3\nu \rangle (\langle 13 \rangle [13]+\langle 23 \rangle [23])}(2 \langle 3\nu \rangle^3 [12] [13] [23]^2 \langle 12 \rangle^4+\langle 3\nu \rangle^2 [23] (2 \langle 23 \rangle (\langle 3\nu \rangle [13]-\langle 2\nu \rangle [12]) [23]^2\\
&&+\langle 13 \rangle [13] (-4 \langle 1\nu \rangle [12] [13]+6 \langle 3\nu \rangle [23] [13]-9 \langle 2\nu \rangle [12] [23])) \langle 12 \rangle^3+\langle 3\nu \rangle (-2 \langle 23 \rangle^2 \langle 2\nu \rangle \langle 3\nu \rangle [23]^4+\langle 13 \rangle \langle 23 \rangle \langle 2\nu \rangle (3 \langle 2\nu \rangle [12]\\
&&-11 \langle 3\nu \rangle [13]) [23]^3+\langle 13 \rangle^2 [13] (\langle 1\nu \rangle^2 [12] [13]^2+\langle 1\nu \rangle (5 \langle 2\nu \rangle [12]-4 \langle 3\nu \rangle [13]) [23] [13]+\langle 2\nu \rangle (7 \langle 2\nu \rangle [12]-15 \langle 3\nu \rangle [13]) [23]^2)) \langle 12 \rangle^2\\
&&+\langle 13 \rangle \langle 2\nu \rangle \langle 3\nu \rangle [23] (3 \langle 23 \rangle^2 \langle 2\nu \rangle [23]^3+7 \langle 13 \rangle \langle 23 \rangle \langle 2\nu \rangle [13] [23]^2+\langle 13 \rangle^2 [13]^2 (\langle 1\nu \rangle [13]+3 \langle 2\nu \rangle [23])) \langle 12 \rangle\\
&&+\langle 13 \rangle^2 \langle 2\nu \rangle (\langle 23 \rangle^2 \langle 2\nu \rangle^2 [23]^4+4 \langle 13 \rangle \langle 23 \rangle \langle 2\nu \rangle^2 [13] [23]^3+\langle 13 \rangle^2 [13]^2 (\langle 1\nu \rangle^2 [13]^2+4 \langle 1\nu \rangle \langle 2\nu \rangle [23] [13]+6 \langle 2\nu \rangle^2 [23]^2))) ,\\
&&\\
&&(A^l_\mu(1^+,2^+,3^+) \epsilon_3^\mu)_{C0[1,2,3]}\\
&&=-\frac{[12] (\langle 1\nu \rangle [13]+\langle 2\nu \rangle [23])}{64 \langle 13 \rangle^2 \langle 1\nu \rangle \langle 23 \rangle \langle 2\nu \rangle \langle 3\nu \rangle^2}(\langle 2\nu \rangle (2 \langle 1\nu \rangle \langle 2\nu \rangle [12]+3 \langle 1\nu \rangle \langle 3\nu \rangle [13]+\langle 2\nu \rangle \langle 3\nu \rangle [23]) \langle 13 \rangle^2+\langle 12 \rangle \langle 3\nu \rangle (\langle 1\nu \rangle \langle 2\nu \rangle [12]\\
&&-2 \langle 1\nu \rangle \langle 3\nu \rangle [13]-3 \langle 2\nu \rangle \langle 3\nu \rangle [23]) \langle 13 \rangle+2 \langle 12 \rangle^2 \langle 3\nu \rangle^2 (\langle 3\nu \rangle [23]-\langle 1\nu \rangle [12])) ,\\
&&\\
&&(A^l_\mu(1^+,2^+,3^+) \epsilon_3^\mu)_{B0[2,4]}\\
&&=\frac{(\langle 1\nu \rangle [13]+\langle 2\nu \rangle [23])}{64 \langle 12 \rangle \langle 13 \rangle \langle 1\nu \rangle \langle 23 \rangle \langle 2\nu \rangle \langle 3\nu \rangle^2 (\langle 12 \rangle [12]+\langle 13 \rangle [13])^2} (4 \langle 3\nu \rangle^2 [12] [13] (\langle 1\nu \rangle [12]-\langle 3\nu \rangle [23]) \langle 12 \rangle^3+\langle 13 \rangle (\langle 1\nu \rangle [12] (2 \langle 2\nu \rangle^2 [12]^2\\
&&-6 \langle 2\nu \rangle \langle 3\nu \rangle [13] [12]+5 \langle 3\nu \rangle^2 [13]^2)+\langle 3\nu \rangle (-4 \langle 2\nu \rangle^2 [12]^2+7 \langle 2\nu \rangle \langle 3\nu \rangle [13] [12]-2 \langle 3\nu \rangle^2 [13]^2) [23]) \langle 12 \rangle^2+\langle 13 \rangle^2 (\langle 1\nu \rangle [13] (7 \langle 2\nu \rangle^2 [12]^2\\
&&-6 \langle 2\nu \rangle \langle 3\nu \rangle [13] [12]+\langle 3\nu \rangle^2 [13]^2)+\langle 2\nu \rangle (3 \langle 2\nu \rangle^2 [12]^2-10 \langle 2\nu \rangle \langle 3\nu \rangle [13] [12]+\langle 3\nu \rangle^2 [13]^2) [23]) \langle 12 \rangle\\
&&+5 \langle 13 \rangle^3 \langle 2\nu \rangle^2 [12] [13] (\langle 1\nu \rangle [13]+\langle 2\nu \rangle [23])) ,\\
&&\\
&&(A^l_\mu(1^+,2^+,3^+) \epsilon_3^\mu)_{B0[1,4]}\\
&&=-\frac{1}{64 \langle 12 \rangle \langle 13 \rangle \langle 1\nu \rangle \langle 23 \rangle \langle 2\nu \rangle \langle 3\nu \rangle^2 (\langle 12 \rangle [12]+\langle 13 \rangle [13])^2 (\langle 13 \rangle [13]+\langle 23 \rangle [23])^2}(-4 \langle 3\nu \rangle^3 [12]^2 [13] [23] (\langle 1\nu \rangle [12] [13]-2 \langle 3\nu \rangle [23] [13]\\
&&+\langle 2\nu \rangle [12] [23]) \langle 12 \rangle^5+\langle 3\nu \rangle^2 [12] [13] (4 \langle 23 \rangle \langle 3\nu \rangle (\langle 3\nu \rangle [13]-2 \langle 2\nu \rangle [12]) [23]^3+\langle 13 \rangle (2 \langle 1\nu \rangle^2 [12]^2 [13]^2+\langle 1\nu \rangle [12] (7 \langle 2\nu \rangle [12]\\
&&-20 \langle 3\nu \rangle [13]) [23] [13]+(9 \langle 2\nu \rangle^2 [12]^2-40 \langle 2\nu \rangle \langle 3\nu \rangle [13] [12]+20 \langle 3\nu \rangle^2 [13]^2) [23]^2)) \langle 12 \rangle^4+\langle 3\nu \rangle (-4 \langle 23 \rangle^2 \langle 2\nu \rangle \langle 3\nu \rangle^2 [12] [13] [23]^4\\
&&+\langle 13 \rangle \langle 23 \rangle (-4 \langle 2\nu \rangle^3 [12]^3+16 \langle 2\nu \rangle^2 \langle 3\nu \rangle [13] [12]^2-31 \langle 2\nu \rangle \langle 3\nu \rangle^2 [13]^2 [12]+2 \langle 3\nu \rangle^3 [13]^3) [23]^3+\langle 13 \rangle^2 [13] (\langle 1\nu \rangle^2 [12]^2 (8 \langle 3\nu \rangle [13]\\
&&-3 \langle 2\nu \rangle [12]) [13]^2+\langle 1\nu \rangle [12] (-11 \langle 2\nu \rangle^2 [12]^2+42 \langle 2\nu \rangle \langle 3\nu \rangle [13] [12]-24 \langle 3\nu \rangle^2 [13]^2) [23] [13]+(-13 \langle 2\nu \rangle^3 [12]^3+74 \langle 2\nu \rangle^2 \langle 3\nu \rangle [13] [12]^2\\
&&-70 \langle 2\nu \rangle \langle 3\nu \rangle^2 [13]^2 [12]+8 \langle 3\nu \rangle^3 [13]^3) [23]^2)) \langle 12 \rangle^3+\langle 13 \rangle (\langle 23 \rangle^2 \langle 2\nu \rangle \langle 3\nu \rangle (-4 \langle 2\nu \rangle^2 [12]^2+7 \langle 2\nu \rangle \langle 3\nu \rangle [13] [12]-2 \langle 3\nu \rangle^2 [13]^2) [23]^4\\
&&+\langle 13 \rangle \langle 23 \rangle \langle 2\nu \rangle (\langle 2\nu \rangle^3 [12]^3-35 \langle 2\nu \rangle^2 \langle 3\nu \rangle [13] [12]^2+43 \langle 2\nu \rangle \langle 3\nu \rangle^2 [13]^2 [12]-11 \langle 3\nu \rangle^3 [13]^3) [23]^3+\langle 13 \rangle^2 [13] (\langle 1\nu \rangle^2 [12] (\langle 2\nu \rangle^2 [12]^2\\
&&-14 \langle 2\nu \rangle \langle 3\nu \rangle [13] [12]+6 \langle 3\nu \rangle^2 [13]^2) [13]^2+\langle 1\nu \rangle (3 \langle 2\nu \rangle^3 [12]^3-58 \langle 2\nu \rangle^2 \langle 3\nu \rangle [13] [12]^2+42 \langle 2\nu \rangle \langle 3\nu \rangle^2 [13]^2 [12]-6 \langle 3\nu \rangle^3 [13]^3) [23] [13]\\
&&+\langle 2\nu \rangle (3 \langle 2\nu \rangle^3 [12]^3-84 \langle 2\nu \rangle^2 \langle 3\nu \rangle [13] [12]^2+98 \langle 2\nu \rangle \langle 3\nu \rangle^2 [13]^2 [12]-16 \langle 3\nu \rangle^3 [13]^3) [23]^2)) \langle 12 \rangle^2+\langle 13 \rangle^2 \langle 2\nu \rangle (\langle 23 \rangle^2 \langle 2\nu \rangle (3 \langle 2\nu \rangle^2 [12]^2\\
&&-10 \langle 2\nu \rangle \langle 3\nu \rangle [13] [12]+\langle 3\nu \rangle^2 [13]^2) [23]^4+3 \langle 13 \rangle \langle 23 \rangle \langle 2\nu \rangle [13] (5 \langle 2\nu \rangle^2 [12]^2-17 \langle 2\nu \rangle \langle 3\nu \rangle [13] [12]+\langle 3\nu \rangle^2 [13]^2) [23]^3\\
&&+\langle 13 \rangle^2 [13]^2 (\langle 1\nu \rangle^2 [12] (6 \langle 2\nu \rangle [12]-11 \langle 3\nu \rangle [13]) [13]^2+\langle 1\nu \rangle (21 \langle 2\nu \rangle^2 [12]^2-53 \langle 2\nu \rangle \langle 3\nu \rangle [13] [12]+\langle 3\nu \rangle^2 [13]^2) [23] [13]\\
&&+3 \langle 2\nu \rangle (9 \langle 2\nu \rangle^2 [12]^2-31 \langle 2\nu \rangle \langle 3\nu \rangle [13] [12]+\langle 3\nu \rangle^2 [13]^2) [23]^2)) \langle 12 \rangle+5 \langle 13 \rangle^3 \langle 2\nu \rangle^2 [12] [13] (\langle 23 \rangle^2 \langle 2\nu \rangle^2 [23]^4\\
&&+4 \langle 13 \rangle \langle 23 \rangle \langle 2\nu \rangle^2 [13] [23]^3+\langle 13 \rangle^2 [13]^2 (\langle 1\nu \rangle^2 [13]^2+4 \langle 1\nu \rangle \langle 2\nu \rangle [23] [13]+6 \langle 2\nu \rangle^2 [23]^2))) ,\\
&&\\
&&(A^l_\mu(1^+,2^+,3^+) \epsilon_3^\mu)_{B0[1,3]}\\
&&=\frac{1}{64 \langle 12 \rangle \langle 13 \rangle \langle 1\nu \rangle \langle 23 \rangle \langle 2\nu \rangle \langle 3\nu \rangle^2 (\langle 13 \rangle [13]+\langle 23 \rangle [23])^2}(\langle 2\nu \rangle [13] (\langle 1\nu \rangle^2 (\langle 2\nu \rangle [12]-\langle 3\nu \rangle [13]) [13]^2+\langle 1\nu \rangle \langle 2\nu \rangle (3 \langle 2\nu \rangle [12]\\
&&-5 \langle 3\nu \rangle [13]) [23] [13]+3 \langle 2\nu \rangle^2 (\langle 2\nu \rangle [12]-3 \langle 3\nu \rangle [13]) [23]^2) \langle 13 \rangle^3+(\langle 23 \rangle \langle 2\nu \rangle^3 (\langle 2\nu \rangle [12]-7 \langle 3\nu \rangle [13]) [23]^3+\langle 12 \rangle \langle 3\nu \rangle [13] (\langle 1\nu \rangle^2 (\langle 3\nu \rangle [13]\\
&&-3 \langle 2\nu \rangle [12]) [13]^2+\langle 1\nu \rangle \langle 2\nu \rangle (5 \langle 3\nu \rangle [13]-11 \langle 2\nu \rangle [12]) [23] [13]+\langle 2\nu \rangle^2 (11 \langle 3\nu \rangle [13]-13 \langle 2\nu \rangle [12]) [23]^2)) \langle 13 \rangle^2+\langle 3\nu \rangle (-2 \langle 23 \rangle^2 \langle 2\nu \rangle^3 [23]^4\\
&&+4 \langle 12 \rangle \langle 23 \rangle \langle 2\nu \rangle^2 (\langle 3\nu \rangle [13]-\langle 2\nu \rangle [12]) [23]^3+\langle 12 \rangle^2 \langle 3\nu \rangle [13] (2 \langle 1\nu \rangle^2 [12] [13]^2+\langle 1\nu \rangle (7 \langle 2\nu \rangle [12]-5 \langle 3\nu \rangle [13]) [23] [13]\\
&&+\langle 2\nu \rangle (9 \langle 2\nu \rangle [12]-11 \langle 3\nu \rangle [13]) [23]^2)) \langle 13 \rangle-4 \langle 12 \rangle^2 \langle 3\nu \rangle^3 [13] [23] (\langle 23 \rangle \langle 2\nu \rangle [23]^2+\langle 12 \rangle (\langle 1\nu \rangle [12] [13]-\langle 3\nu \rangle [23] [13]+\langle 2\nu \rangle [12] [23]))) .
\eeas
\end{scriptsize}

For $A^l_\mu(1^+,2^+,3^+) k_1^\mu$:
\begin{scriptsize}
\beas
&&(A^l_\mu(1^+,2^+,3^+) k_1^\mu)_{D0[1,2,3,4]}\\
&&=\frac{[12]}{64 \langle 13 \rangle^2 \langle 1\nu \rangle \langle 2\nu \rangle \langle 3\nu \rangle} (\langle 2\nu \rangle (4 \langle 2\nu \rangle [12]+5 \langle 3\nu \rangle [13]) \langle 13 \rangle^2+3 \langle 12 \rangle \langle 2\nu \rangle \langle 3\nu \rangle [12] \langle 13 \rangle-2 \langle 12 \rangle^2 \langle 3\nu \rangle^2 [12]) [23] (\langle 12 \rangle \langle 1\nu \rangle [12]\\
&&+\langle 13 \rangle \langle 1\nu \rangle [13]+\langle 13 \rangle \langle 2\nu \rangle [23]-\langle 12 \rangle \langle 3\nu \rangle [23]),\\
&&\\
&&(A^l_\mu(1^+,2^+,3^+) k_1^\mu)_{C0[1,3,4]}\\
&&=\frac{1}{64 \langle 12 \rangle \langle 13 \rangle^2 \langle 1\nu \rangle \langle 23 \rangle \langle 2\nu \rangle \langle 3\nu \rangle}(2 \langle 3\nu \rangle^2 [12]^2 (\langle 3\nu \rangle [23]-\langle 1\nu \rangle [12]) \langle 12 \rangle^4+\langle 13 \rangle \langle 3\nu \rangle [12] (\langle 1\nu \rangle [12] (3 \langle 2\nu \rangle [12]-4 \langle 3\nu \rangle [13])\\
&&+\langle 3\nu \rangle (2 \langle 3\nu \rangle [13]-5 \langle 2\nu \rangle [12]) [23]) \langle 12 \rangle^3+\langle 13 \rangle^2 ([13] (\langle 3\nu \rangle [13]-8 \langle 2\nu \rangle [12]) [23] \langle 3\nu \rangle^2+\langle 1\nu \rangle [12] (2 \langle 2\nu \rangle [12]-\langle 3\nu \rangle [13]) (\langle 2\nu \rangle [12]\\
&&+4 \langle 3\nu \rangle [13])) \langle 12 \rangle^2+\langle 13 \rangle^3 (\langle 1\nu \rangle [13] (6 \langle 2\nu \rangle^2 [12]^2+9 \langle 2\nu \rangle \langle 3\nu \rangle [13] [12]-2 \langle 3\nu \rangle^2 [13]^2)+\langle 2\nu \rangle (3 \langle 2\nu \rangle^2 [12]^2+2 \langle 2\nu \rangle \langle 3\nu \rangle [13] [12]\\
&&-6 \langle 3\nu \rangle^2 [13]^2) [23]) \langle 12 \rangle+\langle 13 \rangle^4 \langle 2\nu \rangle [13] (4 \langle 2\nu \rangle [12]+5 \langle 3\nu \rangle [13]) (\langle 1\nu \rangle [13]+\langle 2\nu \rangle [23])),\\
&&\\
&&(A^l_\mu(1^+,2^+,3^+) k_1^\mu)_{C0[2,3,4]}\\
&&=\frac{[23]}{64 \langle 12 \rangle \langle 13 \rangle^2 \langle 1\nu \rangle \langle 2\nu \rangle \langle 3\nu \rangle}
 (2 \langle 3\nu \rangle^2 [12] (\langle 1\nu \rangle [12]-\langle 3\nu \rangle [23]) \langle 12 \rangle^3+\langle 13 \rangle \langle 3\nu \rangle [12] (-3 \langle 1\nu \rangle \langle 2\nu \rangle [12]+2 \langle 1\nu \rangle \langle 3\nu \rangle [13]\\
&&+5 \langle 2\nu \rangle \langle 3\nu \rangle [23]) \langle 12 \rangle^2+\langle 13 \rangle^2 \langle 2\nu \rangle (2 \langle 1\nu \rangle [12] (2 \langle 2\nu \rangle [12]+\langle 3\nu \rangle [13])-\langle 3\nu \rangle (9 \langle 2\nu \rangle [12]+7 \langle 3\nu \rangle [13]) [23]) \langle 12 \rangle\\
&&+\langle 13 \rangle^3 \langle 2\nu \rangle (4 \langle 2\nu \rangle [12]+5 \langle 3\nu \rangle [13]) (\langle 1\nu \rangle [13]+\langle 2\nu \rangle [23])),\\
&&\\
&&(A^l_\mu(1^+,2^+,3^+) k_1^\mu)_{C0[1,3,4]}\\
&&=\frac{1}{64 \langle 12 \rangle \langle 13 \rangle^2 \langle 1\nu \rangle \langle 23 \rangle \langle 2\nu \rangle \langle 3\nu \rangle (\langle 13 \rangle [13]+\langle 23 \rangle [23])}(2 \langle 3\nu \rangle^3 [12]^2 [23] (2 \langle 13 \rangle [13]+\langle 23 \rangle [23]) \langle 12 \rangle^4+\langle 3\nu \rangle^2 [12] (2 \langle 23 \rangle^2 \langle 3\nu \rangle [23]^3\\
&&+\langle 13 \rangle \langle 23 \rangle (6 \langle 3\nu \rangle [13]-5 \langle 2\nu \rangle [12]) [23]^2-\langle 13 \rangle^2 [13] (\langle 1\nu \rangle [12] [13]-2 \langle 3\nu \rangle [23] [13]+10 \langle 2\nu \rangle [12] [23])) \langle 12 \rangle^3-\langle 13 \rangle \langle 3\nu \rangle (5 \langle 23 \rangle^2 \langle 2\nu \rangle \langle 3\nu \rangle [12] [23]^3\\
&&+2 \langle 13 \rangle \langle 23 \rangle \langle 3\nu \rangle [13] (9 \langle 2\nu \rangle [12]+2 \langle 3\nu \rangle [13]) [23]^2+\langle 13 \rangle^2 [13] ((2 \langle 2\nu \rangle^2 [12]^2+17 \langle 2\nu \rangle \langle 3\nu \rangle [13] [12]+6 \langle 3\nu \rangle^2 [13]^2) [23]-\langle 1\nu \rangle [12] [13] (2 \langle 2\nu \rangle [12]\\
&&+\langle 3\nu \rangle [13]))) \langle 12 \rangle^2+\langle 13 \rangle^2 (\langle 23 \rangle^2 \langle 2\nu \rangle \langle 3\nu \rangle (\langle 2\nu \rangle [12]-\langle 3\nu \rangle [13]) [23]^3+\langle 13 \rangle \langle 23 \rangle \langle 2\nu \rangle (4 \langle 2\nu \rangle^2 [12]^2+3 \langle 2\nu \rangle \langle 3\nu \rangle [13] [12]-3 \langle 3\nu \rangle^2 [13]^2) [23]^2\\
&&+\langle 13 \rangle^2 [13] (\langle 1\nu \rangle [13] (4 \langle 2\nu \rangle^2 [12]^2+5 \langle 2\nu \rangle \langle 3\nu \rangle [13] [12]+2 \langle 3\nu \rangle^2 [13]^2)+\langle 2\nu \rangle (8 \langle 2\nu \rangle^2 [12]^2+3 \langle 2\nu \rangle \langle 3\nu \rangle [13] [12]-3 \langle 3\nu \rangle^2 [13]^2) [23])) \langle 12 \rangle\\
&&+\langle 13 \rangle^3 \langle 2\nu \rangle (4 \langle 2\nu \rangle [12]+3 \langle 3\nu \rangle [13]) (\langle 23 \rangle^2 \langle 2\nu \rangle [23]^3+3 \langle 13 \rangle \langle 23 \rangle \langle 2\nu \rangle [13] [23]^2+\langle 13 \rangle^2 [13]^2 (\langle 1\nu \rangle [13]+3 \langle 2\nu \rangle [23]))),\\
&&\\
&&(A^l_\mu(1^+,2^+,3^+) k_1^\mu)_{C0[1,2,3]}\\
&&=\frac{[12]}{64 \langle 13 \rangle^2 \langle 1\nu \rangle \langle 23 \rangle \langle 2\nu \rangle \langle 3\nu \rangle} (\langle 1\nu \rangle (\langle 12 \rangle [12]+\langle 13 \rangle [13]) (\langle 2\nu \rangle (6 \langle 2\nu \rangle [12]+5 \langle 3\nu \rangle [13]) \langle 13 \rangle^2-3 \langle 12 \rangle \langle 2\nu \rangle \langle 3\nu \rangle [12] \langle 13 \rangle+2 \langle 12 \rangle^2 \langle 3\nu \rangle^2 [12])\\
&&+(\langle 13 \rangle \langle 2\nu \rangle-\langle 12 \rangle \langle 3\nu \rangle) (\langle 2\nu \rangle (4 \langle 2\nu \rangle [12]+3 \langle 3\nu \rangle [13]) \langle 13 \rangle^2-3 \langle 12 \rangle \langle 2\nu \rangle \langle 3\nu \rangle [12] \langle 13 \rangle+2 \langle 12 \rangle^2 \langle 3\nu \rangle^2 [12]) [23]),\\
&&\\
&&(A^l_\mu(1^+,2^+,3^+) k_1^\mu)_{B0[2,4]}\\
&&=\frac{1}{64 \langle 12 \rangle \langle 13 \rangle \langle 1\nu \rangle \langle 23 \rangle \langle 2\nu \rangle \langle 3\nu \rangle (\langle 12 \rangle [12]+\langle 13 \rangle [13])}(\langle 1\nu \rangle (\langle 12 \rangle [12]+\langle 13 \rangle [13]) (5 \langle 13 \rangle^2 [12] [13] \langle 2\nu \rangle^2+4 \langle 12 \rangle^2 \langle 3\nu \rangle^2 [12] [13]\\
&&+\langle 12 \rangle \langle 13 \rangle (2 \langle 2\nu \rangle^2 [12]^2-6 \langle 2\nu \rangle \langle 3\nu \rangle [13] [12]+\langle 3\nu \rangle^2 [13]^2))+(5 \langle 13 \rangle^3 [12] [13] \langle 2\nu \rangle^3+13 \langle 12 \rangle^2 \langle 13 \rangle \langle 3\nu \rangle^2 [12] [13] \langle 2\nu \rangle\\
&&+\langle 12 \rangle \langle 13 \rangle^2 (\langle 2\nu \rangle^2 [12]^2-12 \langle 2\nu \rangle \langle 3\nu \rangle [13] [12]+\langle 3\nu \rangle^2 [13]^2) \langle 2\nu \rangle-4 \langle 12 \rangle^3 \langle 3\nu \rangle^3 [12] [13]) [23]),\\
&&\\
&&(A^l_\mu(1^+,2^+,3^+) k_1^\mu)_{B0[1,4]}\\
&&=\frac{1}{64 \langle 12 \rangle \langle 13 \rangle \langle 1\nu \rangle \langle 23 \rangle \langle 2\nu \rangle \langle 3\nu \rangle (\langle 12 \rangle [12]+\langle 13 \rangle [13]) (\langle 13 \rangle [13]+\langle 23 \rangle [23])^3}(2 \langle 3\nu \rangle^3 [12]^3 [13] [23] (3 \langle 13 \rangle [13]+2 \langle 23 \rangle [23]) \langle 12 \rangle^5\\
&&+\langle 3\nu \rangle^2 [12]^2 [13] (8 \langle 23 \rangle^2 \langle 3\nu \rangle [23]^3-11 \langle 13 \rangle \langle 23 \rangle (\langle 2\nu \rangle [12]-2 \langle 3\nu \rangle [13]) [23]^2-2 \langle 13 \rangle^2 [13] (\langle 1\nu \rangle [12] [13]-10 \langle 3\nu \rangle [23] [13]+9 \langle 2\nu \rangle [12] [23])) \langle 12 \rangle^4\\
&&+\langle 3\nu \rangle [12] (4 \langle 23 \rangle^3 \langle 3\nu \rangle^2 [13] [23]^4-4 \langle 13 \rangle \langle 23 \rangle^2 (\langle 2\nu \rangle^2 [12]^2+7 \langle 2\nu \rangle \langle 3\nu \rangle [13] [12]-4 \langle 3\nu \rangle^2 [13]^2) [23]^3+2 \langle 13 \rangle^2 \langle 23 \rangle [13] (-2 \langle 2\nu \rangle^2 [12]^2\\
&&-42 \langle 2\nu \rangle \langle 3\nu \rangle [13] [12]+11 \langle 3\nu \rangle^2 [13]^2) [23]^2+\langle 13 \rangle^3 [13]^2 (\langle 1\nu \rangle [12] [13] (5 \langle 2\nu \rangle [12]-6 \langle 3\nu \rangle [13])+2 (2 \langle 2\nu \rangle^2 [12]^2-40 \langle 2\nu \rangle \langle 3\nu \rangle [13] [12]\\
&&+7 \langle 3\nu \rangle^2 [13]^2) [23])) \langle 12 \rangle^3+\langle 13 \rangle (-\langle 23 \rangle^3 \langle 2\nu \rangle \langle 3\nu \rangle [12] (4 \langle 2\nu \rangle [12]+17 \langle 3\nu \rangle [13]) [23]^4+\langle 13 \rangle \langle 23 \rangle^2 (\langle 2\nu \rangle^3 [12]^3-2 \langle 2\nu \rangle^2 \langle 3\nu \rangle [13] [12]^2\\
&&-83 \langle 2\nu \rangle \langle 3\nu \rangle^2 [13]^2 [12]-4 \langle 3\nu \rangle^3 [13]^3) [23]^3+\langle 13 \rangle^2 \langle 23 \rangle [13] (3 \langle 2\nu \rangle^3 [12]^3+26 \langle 2\nu \rangle^2 \langle 3\nu \rangle [13] [12]^2-136 \langle 2\nu \rangle \langle 3\nu \rangle^2 [13]^2 [12]\\
&&-8 \langle 3\nu \rangle^3 [13]^3) [23]^2+\langle 13 \rangle^3 [13]^2 (\langle 1\nu \rangle [12] [13] (\langle 2\nu \rangle^2 [12]^2+18 \langle 2\nu \rangle \langle 3\nu \rangle [13] [12]-4 \langle 3\nu \rangle^2 [13]^2)+(3 \langle 2\nu \rangle^3 [12]^3+42 \langle 2\nu \rangle^2 \langle 3\nu \rangle [13] [12]^2\\
&&-91 \langle 2\nu \rangle \langle 3\nu \rangle^2 [13]^2 [12]-4 \langle 3\nu \rangle^3 [13]^3) [23])) \langle 12 \rangle^2+\langle 13 \rangle^2 \langle 2\nu \rangle (\langle 23 \rangle^3 (3 \langle 2\nu \rangle^2 [12]^2+12 \langle 2\nu \rangle \langle 3\nu \rangle [13] [12]-5 \langle 3\nu \rangle^2 [13]^2) [23]^4\\
&&+\langle 13 \rangle \langle 23 \rangle^2 [13] (3 \langle 2\nu \rangle [12]-\langle 3\nu \rangle [13]) (3 \langle 2\nu \rangle [12]+19 \langle 3\nu \rangle [13]) [23]^3+9 \langle 13 \rangle^2 \langle 23 \rangle [13]^2 (\langle 2\nu \rangle^2 [12]^2+10 \langle 2\nu \rangle \langle 3\nu \rangle [13] [12]\\
&&-3 \langle 3\nu \rangle^2 [13]^2) [23]^2+\langle 13 \rangle^3 [13]^3 (17 \langle 1\nu \rangle \langle 3\nu \rangle [12] [13]^2+(3 \langle 2\nu \rangle^2 [12]^2+66 \langle 2\nu \rangle \langle 3\nu \rangle [13] [12]-17 \langle 3\nu \rangle^2 [13]^2) [23])) \langle 12 \rangle\\
&&-\langle 13 \rangle^3 \langle 2\nu \rangle [13] (\langle 2\nu \rangle [12]-4 \langle 3\nu \rangle [13]) (\langle 23 \rangle^3 \langle 2\nu \rangle [23]^4+4 \langle 13 \rangle \langle 23 \rangle^2 \langle 2\nu \rangle [13] [23]^3+6 \langle 13 \rangle^2 \langle 23 \rangle \langle 2\nu \rangle [13]^2 [23]^2\\
&&+\langle 13 \rangle^3 [13]^3 (\langle 1\nu \rangle [13]+4 \langle 2\nu \rangle [23]))),\\
&&\\
&&(A^l_\mu(1^+,2^+,3^+) k_1^\mu)_{B0[1,3]}\\
&&=-\frac{1}{64 \langle 12 \rangle \langle 13 \rangle \langle 1\nu \rangle \langle 23 \rangle \langle 2\nu \rangle \langle 3\nu \rangle (\langle 13 \rangle [13]+\langle 23 \rangle [23])^3}(2 \langle 3\nu \rangle^3 [12]^2 [13] [23] (3 \langle 13 \rangle [13]+2 \langle 23 \rangle [23]) \langle 12 \rangle^4+\langle 3\nu \rangle^2 [12] [13] (4 \langle 23 \rangle^2 \langle 3\nu \rangle [23]^3\\
&&+\langle 13 \rangle \langle 23 \rangle (7 \langle 3\nu \rangle [13]-11 \langle 2\nu \rangle [12]) [23]^2-2 \langle 13 \rangle^2 [13] (\langle 1\nu \rangle [12] [13]-2 \langle 3\nu \rangle [23] [13]+9 \langle 2\nu \rangle [12] [23])) \langle 12 \rangle^3\\
&&-\langle 13 \rangle \langle 3\nu \rangle [13] (2 \langle 23 \rangle^2 \langle 3\nu \rangle (7 \langle 2\nu \rangle [12]+2 \langle 3\nu \rangle [13]) [23]^3+\langle 13 \rangle \langle 23 \rangle (-8 \langle 2\nu \rangle^2 [12]^2+33 \langle 2\nu \rangle \langle 3\nu \rangle [13] [12]+9 \langle 3\nu \rangle^2 [13]^2) [23]^2\\
&&+\langle 13 \rangle^2 [13] (\langle 1\nu \rangle [12] [13] (\langle 3\nu \rangle [13]-5 \langle 2\nu \rangle [12])+2 (-8 \langle 2\nu \rangle^2 [12]^2+12 \langle 2\nu \rangle \langle 3\nu \rangle [13] [12]+3 \langle 3\nu \rangle^2 [13]^2) [23])) \langle 12 \rangle^2\\
&&+\langle 13 \rangle (2 \langle 23 \rangle^3 \langle 2\nu \rangle^2 \langle 3\nu \rangle [12] [23]^4+\langle 13 \rangle \langle 23 \rangle^2 \langle 2\nu \rangle (-3 \langle 2\nu \rangle^2 [12]^2+13 \langle 2\nu \rangle \langle 3\nu \rangle [13] [12]+2 \langle 3\nu \rangle^2 [13]^2) [23]^3\\
&&+3 \langle 13 \rangle^2 \langle 23 \rangle \langle 2\nu \rangle [13] (-3 \langle 2\nu \rangle^2 [12]^2+9 \langle 2\nu \rangle \langle 3\nu \rangle [13] [12]+2 \langle 3\nu \rangle^2 [13]^2) [23]^2+\langle 13 \rangle^3 [13]^2 (\langle 1\nu \rangle [13] (-3 \langle 2\nu \rangle^2 [12]^2+4 \langle 2\nu \rangle \langle 3\nu \rangle [13] [12]\\
&&+\langle 3\nu \rangle^2 [13]^2)+\langle 2\nu \rangle (-9 \langle 2\nu \rangle^2 [12]^2+23 \langle 2\nu \rangle \langle 3\nu \rangle [13] [12]+6 \langle 3\nu \rangle^2 [13]^2) [23])) \langle 12 \rangle-\langle 13 \rangle^2 \langle 2\nu \rangle (2 \langle 23 \rangle^3 \langle 2\nu \rangle (2 \langle 2\nu \rangle [12]\\
&&+\langle 3\nu \rangle [13]) [23]^4+\langle 13 \rangle \langle 23 \rangle^2 \langle 2\nu \rangle [13] (15 \langle 2\nu \rangle [12]+7 \langle 3\nu \rangle [13]) [23]^3+3 \langle 13 \rangle^2 \langle 23 \rangle \langle 2\nu \rangle [13]^2 (7 \langle 2\nu \rangle [12]+3 \langle 3\nu \rangle [13]) [23]^2\\
&&+\langle 13 \rangle^3 [13]^3 (\langle 1\nu \rangle [13] (3 \langle 2\nu \rangle [12]+\langle 3\nu \rangle [13])+\langle 2\nu \rangle (13 \langle 2\nu \rangle [12]+5 \langle 3\nu \rangle [13]) [23]))).
\eeas
\end{scriptsize}

We checked our four point amplitudes using the known simple results of $A^l(1^+,2^+,3^+,4^+)$ and $A^l(1^+,2^+,3^+,4^-)$ in \cite{Bern:1991aq,Bern2,Kharel,Mahlon:1993si}.

%the reader can ask us for full four point amplitudes

%\textcolor{red}{this expression in some specific Lorentz frame?}

%Due to the length of the expression, we put the result in the Appendix. In the Appendix we only show the result with the other three on shell legs to be of positive helicity ie. $A^{l\ +++}_\mu$, although we have calculated all other helicity results. The length of the expression makes it seem formidable to calculate the analytic expressions of more point off shell amplitudes, where numerical calculations should replace the analytic ones. 

%We have checked our result of $A^l_\mu(k_1,k_2,k_3)$ by some well known and simple results like $A^l(1^+, 2^+, 3^+, 4^+)$ and $A^l(1^+, 2^+, 3^+, 4^-)$ etc.\textcolor{red}{citations} and also by Ward identity.

\section{Conclusion}
We have discussed the Ward identity in detail for off shell amplitudes in pure Yang-Mills theory.  We explicitly prove that the Ward identity with two complexified external lines holds at tree and one loop level using Feynman rules. Then we use the Ward identity to deduce recursion relations for off shell amplitudes at tree and one loop level. In this technique, three steps are important to simplify the calculation. First, according to the complexfied Ward identity, we can convert the calculation of the amplitudes to the calculation of derivative of the amplitudes. Second, we decompose the three point vertex which contains the off shell line into three terms, which simplifies many steps in our calculation. Thirdly, according to the cancellation details in the proof of complexified Ward identity, we find most terms from different diagrams cancel with each other. The number of remaining effective terms or diagrams are reduced. It turns out that the recursion relation we derive at tree level is equivalent to Berends-Giele recursion relation \cite{Berends:1987me}. However, our expressions at 1-loop level are new. And we present 1-loop off shell three and four point amplitudes as examples of applying our method at 1-loop level.

%\textcolor{red}{Second, we decompose the three point vertex which contains the off shell line into three terms, such that when the vertex is on the loop only two terms contribute, and otherwise only one term is needed.}

%\textcolor{red}{We give a full expression of four point one loop off shell amplitudes in the appendix. The length of the expression forbids us to calculate and give more analytic expressions for more point off shell amplitudes. Numerical calculations of one loop off shell amplitudes should be developed for calculations relevant for collider physics etc..}

Comparing with our previous work \cite{Chen}, we find the technique in this article is more universal. Here we can obtain a recursion relation for the total amplitudes instead of just the boundary terms of the amplitudes, and we do not need to use BCFW recursion relation. Furthermore,  for this technique, we do not need to avoid the unphysical poles from the polarization vectors of the shifted on shell leg which can also depend on $z$. Hence this technique works well for the amplitudes with any helicity structure and the momenta shifts are more general than the ones in \cite{Chen}. In addition, this technique can be used for calculating one loop off shell amplitudes with any helicity structure. 

In principle, it is possible to generalize our method to higher loop levels and to other theories such as QCD.  The only obstruct is to classify all the cancellation details for the Ward identity with complexfied external momenta. We leave this to future work. Another extension is to combine our technique with other methods, such as unitary cut, generalized unitary cut, BCFW, OPP \cite{OPP}  etc. to further simplify the calculation in pure Yang-Mills theory.  

%why only OPP citation is given.

\textbf{Acknowledgement} We thank Edna Cheung, Jens Fjelstad, Konstantin G. Savvidy for helpful discussions.  This work is funded by the Priority Academic Program Development of Jiangsu Higher Education Institutions (PAPD), NSFC grant No.~10775067, Research Links Programme of Swedish Research Council under contract No.~348-2008-6049, the Chinese Central Government's 985 Project grants for Nanjing University, the China Science Postdoc grant no. 020400383. the postdoc grants of Nanjing University 0201003020

%\textcolor{red}{NITS}

%\section*{\center Appendix}

%In this appendix, we will present $A^l_\mu(1^+,2^+,3^+)$. We use the integral reduction method in \cite{Ellis} to reduce the integrals to scalar integrals. We use the following notations for the scalar integrations:

%Other scalar integrals for four point amplitude vanish when $k_1, k_2, k_3$ are on shell. The evaluation of these integrals see \cite{BernD2}.

%By the way, we first give one loop two point function:(coupling constants not included)

%and three point one loop off shell amplitude:

%At four point, the length of the expression grows very quickly, and we will only give $A^l_\mu(1^+,2^+,3^+)$. Instead of giving this expression directly, we will give $A^l_\mu(1^+,2^+,3^+) \epsilon_1^\mu$, $A^l_\mu(1^+,2^+,3^+) \epsilon_2^\mu$ and $A^l_\mu(1^+,2^+,3^+) k_1^\mu$. Together with $A^l_\mu(1^+,2^+,3^+) (k_1+k_2+k_3)^\mu=0$, the expressions are enough to determine all the components of $A^l_\mu(1^+,2^+,3^+)$. We choose the spinor representations for $k_{1,2,3}$ and $\epsilon_{1,2,3}$ to be:

%with $\lambda_\nu$ an arbitrary reference spinor. We will use $<\!\!\nu 1\!\!>$ to stand for $<\!\!\lambda_\nu \lambda_1\!\!>$ and similarly others. We use $(A\cdot \epsilon_1)_{D0[1,2,3,4]}$ to denote for the coefficient of $D0[1,2,3,4]$ in $A^l_\mu(1^+,2^+,3^+) \epsilon_1^\mu$, and similarly for others.

%Then for $A^l_\mu(1^+,2^+,3^+) \epsilon_1^\mu$:

%\textcolor{red}{this expression in some specific Lorentz frame?}

\end{document}